\documentclass[apj,twocolumn,twocolappendix,numberedappendix]{openjournal}

\usepackage{amsmath}
\usepackage{booktabs}
\usepackage{multirow}
\usepackage{color}
\usepackage{soul}
\usepackage{threeparttable}
\usepackage{float}
\usepackage{graphicx}
\usepackage{CJK}
\usepackage{xspace}
\usepackage{afterpage}
\usepackage{placeins}
\usepackage[breaklinks,colorlinks,citecolor=blue,urlcolor=blue,linkcolor=blue,filecolor=blue]{hyperref}

\usepackage[export]{adjustbox}

\def\farcs{%
 \mbox{%
  \kern  0.13ex.%
  \kern -0.95ex\arcsec%
  \kern -0.1ex%
 }%
}%

\newcommand{\oiii}{[O\,{\sc iii}]}
\newcommand{\heii}{[He\,{\sc ii}]}

\newcommand{\hst}{{\it HST}}

\newcommand{\jwst}{{\it JWST}}

\newcommand{\targ}{GLIMPSE-16043}

\newcommand{\targc}{AMORE6}
\newcommand{\targd}{A370-z6LAE-1}
\newcommand{\targdc}{A370-z6LAE-1c}
\newcommand{\targe}{A370-z6LAE-2}

\shorttitle{
An Exotic Balmer-Jump Galaxy at $z=6.20$}
\shortauthors{Fujimoto, Asada, Naidu et al.}

\begin{document}

\title{
GLIMPSE-\textit{D}: An Exotic Balmer-Jump Object at $z=6.20$?\\
Revisiting Photometric Selection and the Cosmic Abundance of Pop~III Galaxies
\vspace{-15mm}}

\author{
Seiji Fujimoto$^{1,2}$\footnotemark[*],
Yoshihisa Asada$^{1,2,3}$,
Rohan P.\ Naidu$^{4}$\footnotemark[$\dagger$],
John Chisholm$^{5,6}$,
Hakim Atek$^{7}$,
Gabriel Brammer$^{8}$,
Danielle A.\ Berg$^{5}$,
Daniel Schaerer$^{9,10}$,
Vasily Kokorev$^{5,6}$,
Lukas J.\ Furtak$^{5,6}$,
Johan Richard$^{11}$,
Alessandra Venditti$^{12,6}$,
Volker Bromm$^{5,6,13}$,
Angela Adamo$^{14}$,
Adélaïde Claeyssens$^{14}$,
Miroslava Dessauges-Zavadsky$^{9}$,
Qinyue Fei$^{1}$,
Tiger Yu-Yang Hsiao$^{5,6}$,
Damien Korber$^{9}$,
Julian B.\ Mu\~noz$^{5}$,
Richard Pan$^{15}$, and
Alberto Saldana-Lopez$^{14}$
}

\affiliation{$^{1}$David A. Dunlap Department of Astronomy and Astrophysics, University of Toronto, Toronto, ON M5S 3H4, Canada}
\affiliation{$^{2}$Dunlap Institute for Astronomy and Astrophysics, University of Toronto, Toronto, ON M5S 3H4, Canada}
\affiliation{$^{3}$Waseda Research Institute for Science and Engineering, Waseda University, Tokyo 169-8555, Japan}
\affiliation{$^{4}$MIT Kavli Institute for Astrophysics and Space Research, Cambridge, MA 02139, USA}
\affiliation{$^{5}$Department of Astronomy, The University of Texas at Austin, Austin, TX 78712, USA}
\affiliation{$^{6}$Cosmic Frontier Center, The University of Texas at Austin, Austin, TX 78712, USA}
\affiliation{$^{7}$Institut d'Astrophysique de Paris, UMR 7095, CNRS, Sorbonne Université, 98 bis boulevard Arago, 75014 Paris, France}
\affiliation{$^{8}$Cosmic Dawn Center (DAWN), Niels Bohr Institute, University of Copenhagen, DK-2200 Copenhagen, Denmark}
\affiliation{$^{9}$Observatoire de Genève, Université de Genève, Chemin Pegasi 51, 1290 Versoix, Switzerland}
\affiliation{$^{10}$CNRS, IRAP, 14 Avenue E. Belin, 31400 Toulouse, France}
\affiliation{$^{11}$Univ Lyon, Univ Lyon1, Ens de Lyon, CNRS, Centre de Recherche Astrophysique de Lyon UMR5574, 69230 Saint-Genis-Laval, France}
\affiliation{$^{12}$Department of Astronomy, University of Texas at Austin, 2515 Speedway, Stop C1400, Austin, TX 78712, USA}
\affiliation{$^{13}$Weinberg Institute for Theoretical Physics, University of Texas at Austin, Austin, TX 78712, USA}
\affiliation{$^{14}$Department of Astronomy, Oskar Klein Centre, Stockholm University, 106 91 Stockholm, Sweden}
\affiliation{$^{15}$Department of Physics \& Astronomy, Tufts University, MA 02155, USA}

\footnotetext[*]{Email: seiji.fujimoto@utoronto.ca}
\footnotetext[$\dagger$]{Hubble Fellow}

\def\apj{ApJ}%
\def\apjl{ApJL}%
\def\apjs{ApJS}%

\def\rme{\rm e}
\def\rmstar{\rm star}
\def\rmFIR{\rm FIR}
\def\itHubble{\it Hubble}
\def\rmyr{\rm yr}

\begin{abstract}
We present deep \jwst/NIRSpec G395M spectroscopy of \targ, a promising $z \sim 6$ Pop~III candidate originally identified through NIRCam photometry as having weak \oiii$\lambda\lambda4959,5007$ emission. 
Our follow-up reveals clear [O\,\textsc{iii}] emission, ruling out a genuine zero-metallicity nature. 
However, the combination of the measured line fluxes and photometry indicates that its spectral energy distribution requires an extraordinarily strong Balmer jump ($-1.66 \pm 0.47$ mag) and H$\alpha$ equivalent width ($3750 \pm 1800$\,\AA), features that cannot be reproduced by current stellar+nebular or pure nebular photoionization models.
The only models approaching the observations to almost within $1\sigma$ involve a hot ($T_{\rm eff}\!\simeq\!10^{4.7}$ K) single blackbody embedded in a low-$T_{\rm e}$ nebular environment, suggestive of scenarios such as a tidal-disruption event or a microquasar with strong disk winds. 
This cautions that photometric Pop~III selections are vulnerable to contamination when the rest-frame optical continuum is undetected. 
Motivated by this, we refine the photometric Pop~III selection criteria to exclude the locus of extreme Balmer-jump objects. 
The revised criteria also recover the recently reported spectroscopic candidate \targc, demonstrating that the updated selection preserves sensitivity to genuine Pop~III–like sources while removing key contaminants.
Applying the refined criteria across legacy survey fields and five newly released CANUCS lensing cluster fields, we revisit the Pop~III UV luminosity function and estimate the Pop~III cosmic star-formation rate density to be $\approx[10^{-6}$--$10^{-4}]$~$M_{\odot}$~yr$^{-1}$~cMpc$^{-3}$ at $z \simeq 6$--7, falling in the range of current theoretical predictions. 
\end{abstract}

\section{Introduction}
\label{sec:intro}

The formation of the first stars, known as Population~III (Pop~III), marks a pivotal epoch in cosmic history. Emerging from pristine, metal-free gas, these stars initiated the enrichment of the interstellar medium (ISM) and enabled the subsequent formation of chemically evolved galaxies, including the Milky Way \citep[e.g.,][]{bromm2013, frebel2015, klessen2023}. Their massive nature, strong ionizing spectra, and short lifetimes make Pop~III stars critical to understanding early structure formation, cosmic reionization, and the seeding of the first black holes \citep[e.g.,][]{tumlinson2006, bromm2011, xu2016, inayoshi2020}.

Despite their theoretical importance, Pop~III stars and their host galaxies have remained elusive. Prior to the launch of the \textit{James Webb Space Telescope (JWST)}, searches for Pop~III signatures were limited to rest-UV diagnostics in relatively bright systems \citep[e.g.,][]{nagao2008,sobral2015,shibuya2018}, which may not probe the faintest, most metal-poor regimes where Pop~III formation is expected to occur \citep[e.g.,][]{bromm2013}. The advent of \jwst\ has revolutionized the search by enabling access to ultra-faint galaxies at $z \gtrsim 6$ with sufficient sensitivity to detect rest-frame optical emission lines, such as H$\alpha$ and [O~\textsc{iii}], that are key to inferring metallicity and stellar population properties \citep[e.g.,][]{chemerynska2024b}.

In this context, \citet{fujimoto2025} recently presented a novel photometric selection method to identify Pop~III galaxy candidates at $z\simeq6$--7 using deep \jwst/NIRCam imaging \citep[see also e.g.,][]{nagao2008, inoue2011, zackrisson2011, trussler2023, nishigaki2023}. Their search across $\simeq 500$~arcmin$^2$ of legacy survey data led to the discovery of a highly promising candidate, GLIMPSE-16043 (hereafter \targ), at $z \sim 6$. This object exhibited key signatures of a Pop~III-dominated stellar population based on one of the deepest NIRCam photometry so far obtained: extremely strong H$\alpha$ emission, a pronounced Balmer jump, an absence of detectable [O~\textsc{iii}] emission, a blue UV slope, and an inferred stellar age of $\lesssim 5$~Myr. These properties placed ID16043 well within the fiducial Pop~III selection window and motivated the need for spectroscopic follow-up to confirm its nature.

In this paper, we present the results from a dedicated \jwst/NIRSpec deep follow-up program targeting \targ. The program, dubbed \textbf{GLIMPSE-\textit{D}} (PID 9223; PIs: S.~Fujimoto \& R.~Naidu), obtained deep NIRSpec/MSA G395M observations to directly assess the presence or absence of key diagnostic emission lines at rest-frame optical wavelengths. As we discuss in Section~\ref{sec:analysis}, our spectroscopic analysis concludes that \targ\ exhibits clear [O~\textsc{iii}]$\lambda\lambda4959,5007$ emission, definitively ruling out a zero-metallicity Pop~III origin. Nonetheless, \targ\ turns out to be exceptionally unusual from a different aspect: its Balmer jump and the equivalent widths of the Balmer lines are remarkably high, which cannot be reproduced by current general stellar population and photoionization models. We discuss how such objects may mimic Pop~III photometric colors, and we reassess the robustness and limitations of current photometric selection techniques.

This paper is structured as follows. In Section~\ref{sec:obs}, we describe the observations and data reduction. Section~\ref{sec:analysis} outlines the analysis procedures, including emission line measurements and spectral modeling. Section~\ref{sec:true_picture} presents the main spectroscopic results and investigates the physical origin of \targ. In Section~\ref{sec:revisit}, we re-evaluate the photometric Pop~III selection method in light of these findings and provide revised constraints on the Pop~III UV luminosity function and cosmic star formation rate density. We summarize our conclusions in Section~\ref{sec:summary}.

Throughout this paper, we adopt a flat $\Lambda$CDM cosmology with $\Omega_{\mathrm{m}} = 0.3$, $\Omega_{\Lambda} = 0.7$, $H_0 = 70$~km~s$^{-1}$~Mpc$^{-1}$. Magnitudes are given in the AB system \citep{oke1983}.

\section{Observations and Data Reduction}
\label{sec:obs}

The deep spectroscopic follow-up observations of 
\targ\ were carried out with \textit{JWST}/NIRSpec as part of the Director's Discretionary Time (DDT) program GLIMPSE-\textit{D} (PID 9223; PIs: S.~Fujimoto \& R.P.~Naidu). The observations were designed to confirm the Pop~III nature of this photometrically selected candidate by directly testing for the presence of strong H$\alpha$ and H$\beta$ emission and the absence of significant [O~\textsc{iii}]$\lambda\lambda4959,5007$ emission. 
The data were collected using the NIRSpec/Multi-Object Spectroscopy (MOS) mode with the G395M grating and F290LP filter, covering the wavelength range $2.9\,\mu$m–$5.2\,\mu$m at $R\sim1000$ spectral resolution. 

Three separate micro-shutter array (MSA) configurations were used to maximize the number of filler targets while acquiring the full exposure for the main target. These configurations are referred to internally as \texttt{glimpse\_02}, 
\texttt{glimpse\_01}, 
and \texttt{glimpse\_01b}, 
each observed for 42,016~sec, 33,613~sec, and 33,613~sec, respectively. The observations were conducted in two visits: one mask on UT 2025 June 30 and two masks on UT 2025 July 2, resulting in a total on-source integration time of 109,242~s ($\simeq 30.4$~hr) with a consistent MSA position angle. 
To fully utilize the multiplexing capability of the MSA, additional filler slits were assigned to galaxies with photometric redshifts $z_{\mathrm{phot}} > 4$, enabling ancillary science on faint, lensed galaxies in the Abell~S1063 field \citep[see also, e.g.,][]{fei2025,kokorev2025b,berg2025}.

\begin{figure*}[t!]
\begin{center}
\includegraphics[trim=0cm 0cm 0cm 0cm, clip, angle=0,width=1\textwidth]{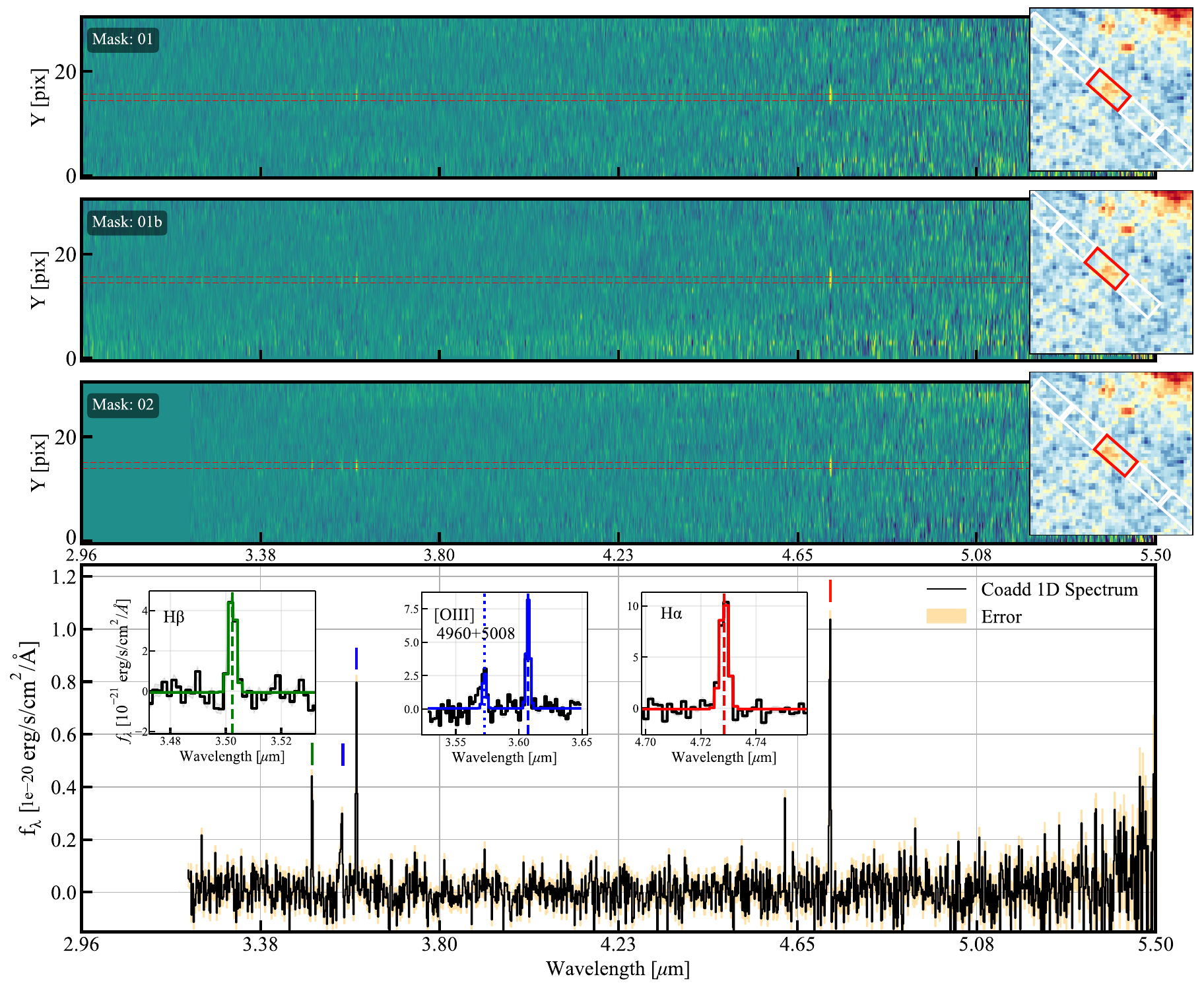}
\end{center}
\vspace{-0.4cm}
 \caption{
NIRSpec/G395M observations of \targ. The top panel shows the reduced 2D spectra from the three MSA configurations, overlaid on the NIRCam/F444W cutout, with the slit positions indicated. The bottom panel presents the co-added 1D spectrum. 
Key optical emission lines of H$\alpha$, [O \textsc{iii}]$\lambda\lambda$4960,5008, and H$\beta$ are all clearly detected. Insets show the best-fit Gaussian profiles to each line with constant continuum terms The measured line fluxes, S/N values, and flux ratios are summarized in Table~\ref{tab:lineflux}.
\label{fig:spectrum}}
\vspace{0.2cm}
\end{figure*}

The data reduction was performed using the \texttt{msaexp} pipeline \citep{brammer2023}, following the procedures described in \citet{heintz2024} and \citet{degraaff2025a}. The reduction methodology is identical to that adopted in \citet{naidu2025} and in the upcoming v4 release of the DAWN \jwst\ NIRSpec archive (Pollock et al., in prep.), where the flux calibration has been updated.
In Figure~\ref{fig:spectrum}, we show the reduced 2D spectra obtained in the three exposures and the shutter configurations around \targ\ overlaid on the NIRCam F444W image cutout. 

\section{Data analysis}
\label{sec:analysis}

\subsection{Co-adding Spectra}
\label{sec:co-add}

The NIRSpec/G395M observations of GLIMPSE-16043 were obtained in three separate MSA configurations. We first scaled each spectrum to a common absolute flux using the NIRCam F444W photometry as a reference, as this band is well within the G395M wavelength coverage and detected with the best S/N owing to the flux boosting of H$\alpha$ line. The synthetic F444W photometry is computed from each reduced spectrum with the F444W filter response curve, and compared to the measured F444W photometric flux to derive a scaling factor. The scaled spectra are then interpolated onto a common wavelength grid and co-added using weights proportional to the H$\alpha$ S/N in each configuration. The co-added flux is taken as the weighted mean of the individual spectra, with uncertainties from the inverse-variance sum. In the bottom panel of Figure~\ref{fig:spectrum}, we present the co-added spectrum, showing the clear detection of H$\alpha$, [O\,\textsc{iii}], and H$\beta$ lines, with no detection of \heii$\lambda$4686. We use this co-added spectrum for all subsequent line measurements throughout this paper.

\subsection{Line Flux Measurements}
\label{sec:line_flux}

We measured the fluxes of the H$\beta$, [O\,\textsc{iii}]~$\lambda\lambda$4960,5008, and H$\alpha$ emission lines from the co-added 1D spectrum by fitting Gaussian profiles. For H$\beta$ and H$\alpha$, we used single-Gaussian models. For [O\,\textsc{iii}], we fit a double-Gaussian model with the flux ratio of $\lambda$5008 to $\lambda$4960 fixed to the theoretical value of 2.98, enforcing a common redshift and width for the two components. The fits were performed in narrow wavelength windows around the expected line centers. Thus, we adopt a constant continuum as a free parameter. Because we find that the redshift forced to be exactly the same number makes the fitting worse, we leave the redshift as a free parameter among these three line fitting. 

In the inset panels at the bottom panel of Figure~\ref{fig:spectrum}, we show the best-fit Gaussian for the three emission lines. The individual line redshifts are consistent within the uncertainties ($<10$~km~s$^{-1}$), and we obtain the best-fit spectroscopic redshift of \targ\ to be 
$z_{\rm spec} = 6.20285 \pm 0.00003$. 
The resulting line fluxes, S/N, and key flux ratios are summarized in Table~\ref{tab:lineflux}. Note that we did not identify the \heii$\lambda$4686 line with $\geq3\sigma$ in the same Gaussian fitting procedure as described above. Using the measurement uncertainty of the H$\beta$ line flux, we also include a 3$\sigma$ upper limit of \heii$\lambda$4686 in Table~\ref{tab:lineflux}. 

\setlength{\tabcolsep}{6pt}
\begin{table}[ht]
\centering
\caption{Emission Line Measurements for GLIMPSE-16043}
\label{tab:lineflux}
\begin{tabular}{lcc}
\hline\hline
Line & S/N & Flux \\
 & & (${\rm erg\,s^{-1}\,cm^{-2}}$)  \\
\hline
\heii$\lambda$4686 & $<3$ & $<1.5\times 10^{-20}$ \\
H$\beta$ & 11.2 & $(1.62 \pm 0.14)\times 10^{-19}$\\
$[$O\,\textsc{iii}$]\,\lambda5008$ & 22.7 & $(2.87 \pm 0.13)\times 10^{-19}$ \\
H$\alpha$ & 23.3 & $(4.18 \pm 0.18)\times 10^{-19}$  \\
\hline
H$\alpha$/H$\beta$ & --- & $2.59 \pm 0.26$  \\
$[$O\,\textsc{iii}$]$/H$\beta$ & --- & $1.78 \pm 0.18$ \\
\hline
\end{tabular}
\end{table}

\section{True Picture of \targ}
\label{sec:true_picture}

\begin{figure*}[t!]
\begin{center}
\includegraphics[width=1.\textwidth]{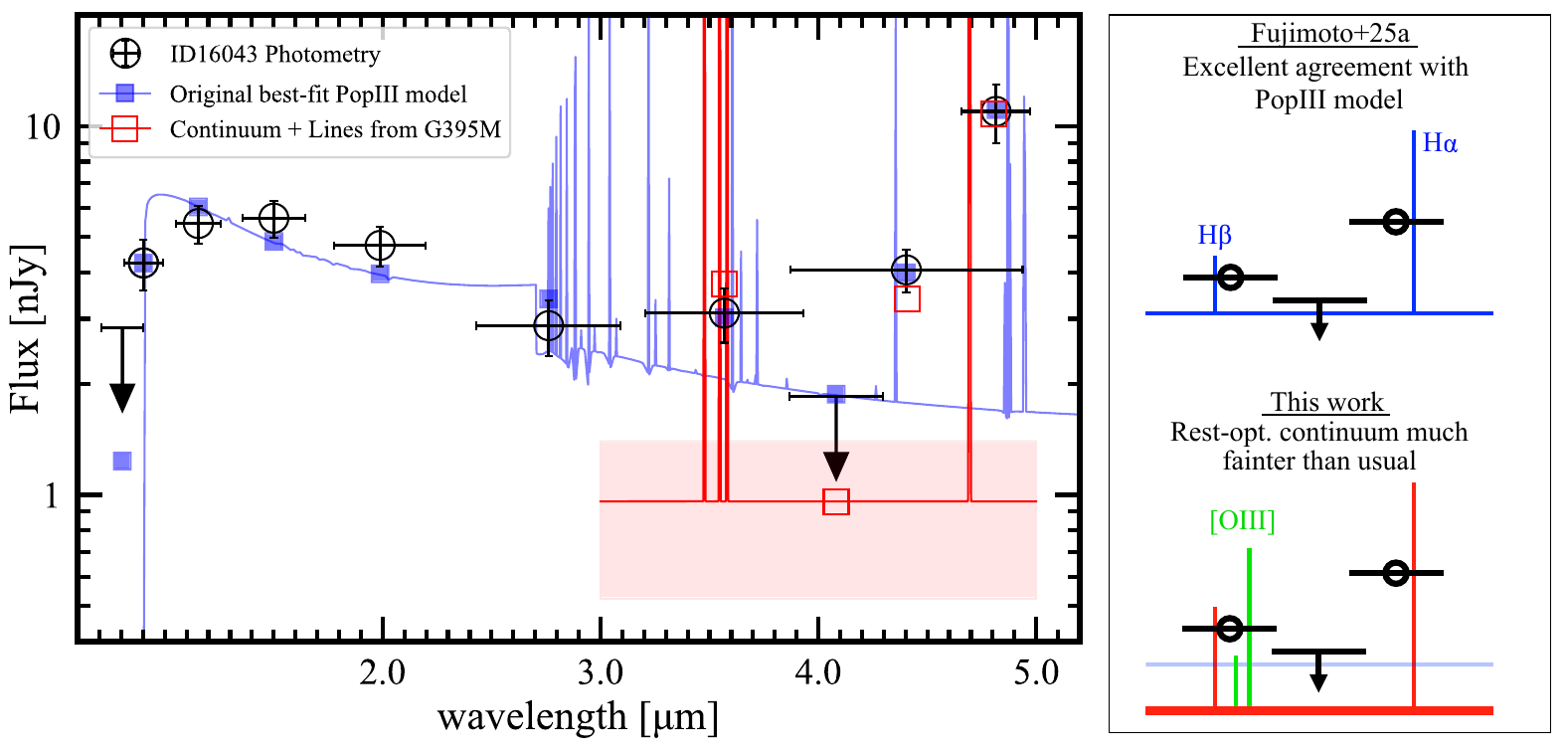}
\end{center}
\vspace{-0.4cm}
\caption{
Comparison of the original Pop~III model fit from \citet{fujimoto2025} and our new best-fit model incorporating the G395M spectroscopic line constraints. 
\textbf{\textit{Left:}}
The red line shows the best-fit model consisting of a flat continuum plus emission lines ([O\,\textsc{iii}] + H$\beta$ and H$\alpha$), with the shaded region indicating the 1$\sigma$ continuum uncertainty. 
Open red squares denote the model-derived synthetic photometry, while black circles with error bars show the observed NIRCam photometry. 
The original Pop~III best-fit model is shown in blue. 
The Pop~III-like photometric feature of the faint F356W excess is no longer prominent once the extremely weak continuum is unveiled, highlighting a cautionary note for photometric Pop~III searches in the absence of a direct continuum detection. 
At the same time, however, the revealed weak continuum results in an extraordinarily strong Balmer jump ($-1.66 \pm 0.47$~mag) and H$\alpha$ equivalent width ($3750 \pm 1800$\,\AA) at $z=6.20$, features that cannot be reproduced by current photoionization models nor by pure nebular emission scenarios (Section~\ref{sec:model}), and this is exactly why the Pop~III solution is favored in \cite{fujimoto2025}. 
\textbf{\textit{Right:}} Illustration summarizing how the unknown underlying continuum leads to the differing interpretations of \targ.
}
\label{fig:cont_fit}
\end{figure*}

\subsection{Possible Causes}
\label{sec:causes}

The clear detection of the [O\,\textsc{iii}] emission line in our G395M spectrum shows that \targ\ is not a truly metal-free (Pop~III) galaxy. In \citet{fujimoto2025}, the classification of \targ\ as a Pop~III candidate was based primarily on the lack of a significant flux excess in the F356W band, where the [O\,\textsc{iii}] line would be expected to fall given its photometric redshift. 
Given the fact that \targ\ turns out not to be a Pop~III galaxy, the possible explanations can be divided into four scenarios:  
(i) an incorrect redshift,  
(ii) photometric uncertainties,  
(iii) non-zero dust attenuation (which would suppress H$\beta$ relative to H$\alpha$), or (iv) an exotic Balmer jump. Below, we explain each scenario and discuss whether it is plausible. 

\textit{(i) Incorrect redshift} -- The Pop~III color–selection method in \citet{fujimoto2025} exploits the flux excesses in F356W and F444W, relative to F410M, arising from [O\,\textsc{iii}]+H$\beta$ and H$\alpha$, respectively. This technique is explicitly valid only for galaxies at $z\simeq5.6$–$6.6$. The spectroscopic redshift from G395M, $z_{\rm spec} = 6.20$, falls well within this range, and the [O\,\textsc{iii}] line is indeed covered by F356W. Thus, an incorrect redshift is not the cause.

\textit{(ii) Photometric uncertainties} -- From the photometric flux-excess measurements, \citet{fujimoto2025} reported a $1\sigma$ upper limit of [O\,\textsc{iii}]/H$\beta < 0.44$. Our G395M measurements yield [O\,\textsc{iii}]/H$\beta = 1.78 \pm 0.18$, a $\sim4\sigma$ discrepancy with respect to the earlier limit. Such a large difference is unlikely to be explained solely by photometric uncertainties.

\textit{(iii) Non-zero dust attenuation} -- In \citet{fujimoto2025}, the contribution of H$\beta$ to the F356W flux excess was subtracted assuming an intrinsic ratio of H$\alpha$/H$\beta = 2.86$ (i.e., case~B recombination without dust), based on the rest-frame UV SED shape. If significant dust attenuation were present, H$\beta$ would be weaker than assumed, increasing the fraction of the F356W excess attributable to [O\,\textsc{iii}]. However, our observed ratio of H$\alpha$/H$\beta = 2.59 \pm 0.26$ is consistent with little to no dust attenuation assuming Case~B recombination, which makes this explanation unlikely.

\textit{(iv) Exotic Balmer jump} -- Measuring the F356W excess requires a reference continuum filter. In the GLIMPSE NIRCam configuration, F410M is expected to sample the underlying continuum of [O\,\textsc{iii}]+H$\beta$ and H$\alpha$ for $z\simeq5.6$–6.6. Notably, \targ\ is undetected in F410M, and in \citet{fujimoto2025} the continuum level was estimated from the best-fit Pop~III SED model to infer the F356W and F444W excesses. If the true continuum is substantially fainter than predicted, the F356W excess would be correspondingly larger, leaving more room for [O\,\textsc{iii}] to contribute. Interestingly, the best-fit Pop~III SED in \citet{fujimoto2025} already features a strong Balmer jump and strong rest-frame H$\alpha$ equivalent width (EW) of $\sim2800\,{\rm \AA}$ to satisfy the stringent F410M limit, and requiring an even fainter continuum implies an unusually large Balmer jump and EW(H$\alpha$) in \targ.

\medskip
In summary, among scenarios (i)–(iv), only (iv) an exceptionally strong Balmer jump remains plausible. In the following, we use the new G395M data to quantify the underlying continuum level, the Balmer jump strength, and the implied nebular conditions in \targ.

\subsection{Measuring Continuum and Balmer Jump Strength}
\label{ref:cont-bj}

In the NIRCam photometry presented in \citet{fujimoto2025}, the rest-frame optical continuum is constrained only by upper limits for \targ. With the G395M spectrum now providing measurements of the key rest-frame optical line fluxes, we can estimate the continuum by subtracting the line contributions from the relevant photometric bands. However, the continuum of \targ\ is extremely faint, and in the flux–scaling procedure described in Section~\ref{sec:co-add}, its contribution to the total photometric flux is negligible. As a result, small uncertainties in the absolute flux scaling could have a disproportionate impact on a direct continuum estimate. To mitigate this, we estimate the underlying continuum using the following approach.

First, we adopt the observed [O\,\textsc{iii}]/H$\beta$ ratio of $1.78 \pm 0.18$, which is minimally impacted by potential uncertainties in the absolute flux calibration. 
Second, we fix the intrinsic case~B Balmer decrement to H$\alpha$/H$\beta = 2.86$ and assume a flat continuum in $f_\nu$. 
We then treat the underlying continuum flux density and the total H$\alpha$ flux as the only free parameters and determine them by fitting the multi-band photometry as described below.

We model the observed flux densities in the F356W, F410M, F444W, and F480M bands as the sum of the flat continuum and the integrated fluxes of the relevant emission lines falling within each filter at $z=6.20$. For F356W, the model includes H$\beta$ and the [O\,\textsc{iii}]~$\lambda\lambda$4960,5008 lines; for F444W and F480M, only H$\alpha$; and for F410M, no strong emission lines. The relative intensities of the [O\,\textsc{iii}] doublet are fixed to the theoretical ratio ($5008/4960 = 2.98$), and H$\beta$ is tied to H$\alpha$ through the assumed Balmer decrement. 
We then calculate the synthetic photometry for the modeled spectrum composed of the underlying continuum and H$\alpha$, H$\beta$, and \oiii\ lines and compare them with the NIRCam photometry. The latest NIRCam photometry for \targ\ is presented in the Appendix \ref{sec:new_photo}. 
Based on the $\chi^2$ minimization approach, we obtain the best-fit continuum flux density of $0.96 \pm 0.44$nJy and H$\alpha$ line flux of $(4.0 \pm 0.60) \times 10^{-19}{\rm erg~s^{-1}cm^{-2}}$. This best-fit H$\alpha$ line flux estimate is in excellent agreement with the direct G395M measurement (Section \ref{sec:line_flux}), providing strong validation for our absolute flux calibration, the emission-line measurements, and the resulting continuum estimate. 

In Figure~\ref{fig:cont_fit}, we present the best-fit model at $\sim$3--5$\mu$m composed of the underlying continuum and emission lines (red curve), together with the corresponding synthetic photometry in the F356W, F410M, F444W, and F480M filters (red square). 
The model-predicted photometry agrees with the observed NIRCam fluxes mostly within the uncertainties, supporting the validity of our continuum and line estimates. 
This result demonstrates that the observed broadband fluxes can be reproduced by a combination of an exceptionally faint continuum and strong emission lines -- including [O\,\textsc{iii}] -- without requiring the purely Pop~III-dominated SED proposed in \citet{fujimoto2025}.

\begin{figure}[t!]
\begin{center}
\includegraphics[width=0.5\textwidth]{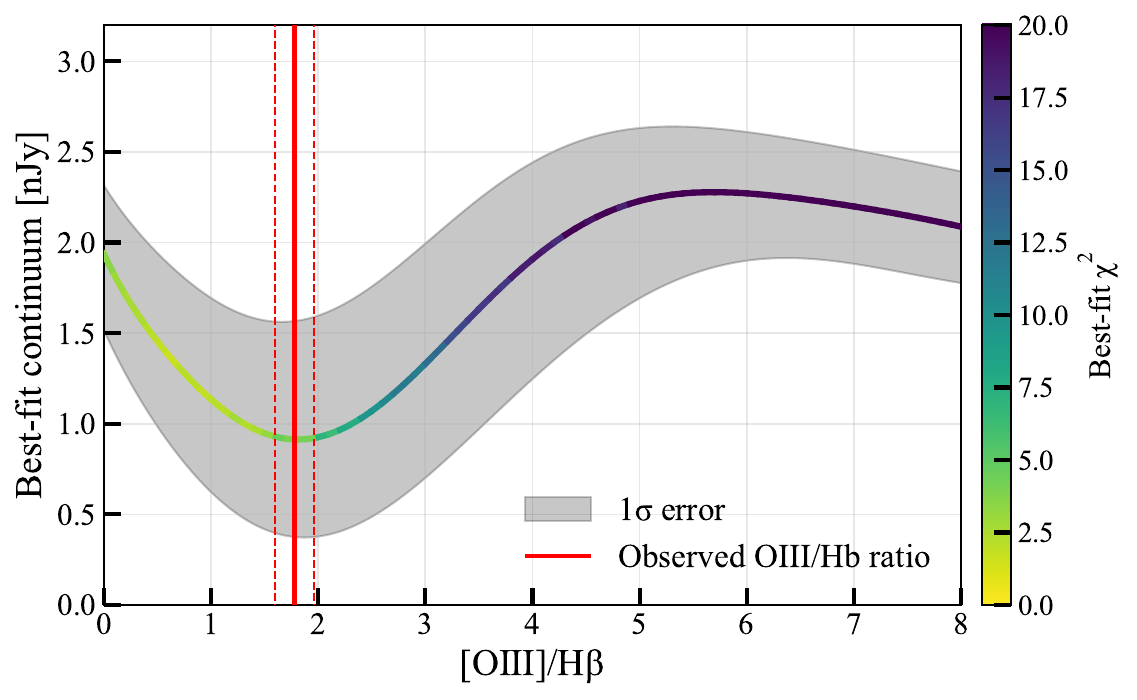}
\end{center}
\vspace{-0.4cm}
\caption{
Relation between the [O\,\textsc{iii}]/H$\beta$ ratio and permitted continuum level from the NIRCam photometry for \targ. The solid curve and shaded area represent the best fit and 1$\sigma$ range at a given [O\,\textsc{iii}]/H$\beta$, color-coded by the $\chi^2$ value from comparing the model synthetic photometry with the observed F356W, F410M, F444W, and F480M fluxes. The vertical solid and dashed red lines denote the observed \oiii/H$\beta$ ratio with NIRSpec/G395M (Section~\ref{sec:line_flux}). 
The curve shape indicates that the inferred very weak continuum and the subsequent extraordinaly strong Balmer jump does not much change from the possible uncertainty of the \oiii/H$\beta$ ratio.  
}
\label{fig:cont-oiiihb}
\end{figure}

In Figure~\ref{fig:cont-oiiihb}, we explore the degeneracy between the continuum level and the assumed [O\,\textsc{iii}]/H$\beta$ ratio in reproducing the observed photometry of \targ. In this test, we repeat the continuum-fitting procedure described above while fixing [O\,\textsc{iii}]/H$\beta$ to values between 0 and 8 in increments of 0.1. For each fixed ratio, we perform the same $\chi^2$ minimization over the continuum and H$\alpha$ flux, and we color-code the results by the resulting $\chi^2$ value, computed using the synthetic photometry from the best-fit model and the observed F356W, F410M, F444W, and F480M fluxes.

Overall, the best-fit continuum level decreases as [O\,\textsc{iii}]/H$\beta$ increases, reflecting the larger fraction of the F356W excess that can be attributed to emission lines. However, for [O\,\textsc{iii}]/H$\beta \gtrsim 2$, the reduced H$\beta$ and H$\alpha$ fluxes require a higher continuum level to match the F444W and F480M photometry. The regime with [O\,\textsc{iii}]/H$\beta > 2$ is disfavored by a rapid increase in $\chi^2$, but the range $0 \lesssim$ [O\,\textsc{iii}]/H$\beta \lesssim 2$ remains highly degenerate if the continuum level is not directly detected. This result cautions against applying photometric Pop~III selection methods when the rest-frame optical continuum is undetected in all relevant filters.

As discussed in scenario~(iv) of Section~\ref{sec:causes}, however, the faint-continuum solution leads to another set of extreme physical properties: an unusually strong Balmer jump and very large EW(H$\alpha$). In fact, our best-fit values imply \textbf{\boldmath EW(H$\alpha$) $= 3750 \pm 1800$~\AA} and a Balmer jump strength, quantified by the difference in AB magnitudes between F200W ($m_{\rm 200}$) and the best-fit optical continuum ($m_{\rm opt,c}$) of \textbf{\boldmath $-1.66 \pm 0.47$~mag}. Albeit the large uncertainty, the EW(H$\alpha$) exceeds the maximum values predicted by stellar+nebular photoionization models even under very top-heavy initial mass functions and/or the presence of very massive stars ($> 100\,M_\odot$) (see e.g., Table~2 in \citealt{schaerer2024}). Likewise, the inferred Balmer jump is nearly twice as large as the values reported in other high-redshift galaxies \citep[e.g.,][]{cameron2024, katz2024}. 

This combination of extreme EW(H$\alpha$) and Balmer jump explains exactly why the \citet{fujimoto2025} analysis favored an [O\,\textsc{iii}]/H$\beta = 0$ solution within the $0$--$2$ range, despite similar $\chi^2$ values from the simple parametric fits (Figure~\ref{fig:cont-oiiihb}), although the large parameter space of metal-enriched galaxies were explored. In the next section, we investigate possible models that can reproduce the remarkably strong EW(H$\alpha$) and Balmer jump observed in \targ

\subsection{Modeling \targ}
\label{sec:model}

\begin{figure*}[t!]
\begin{center}
\includegraphics[width=1.\textwidth]{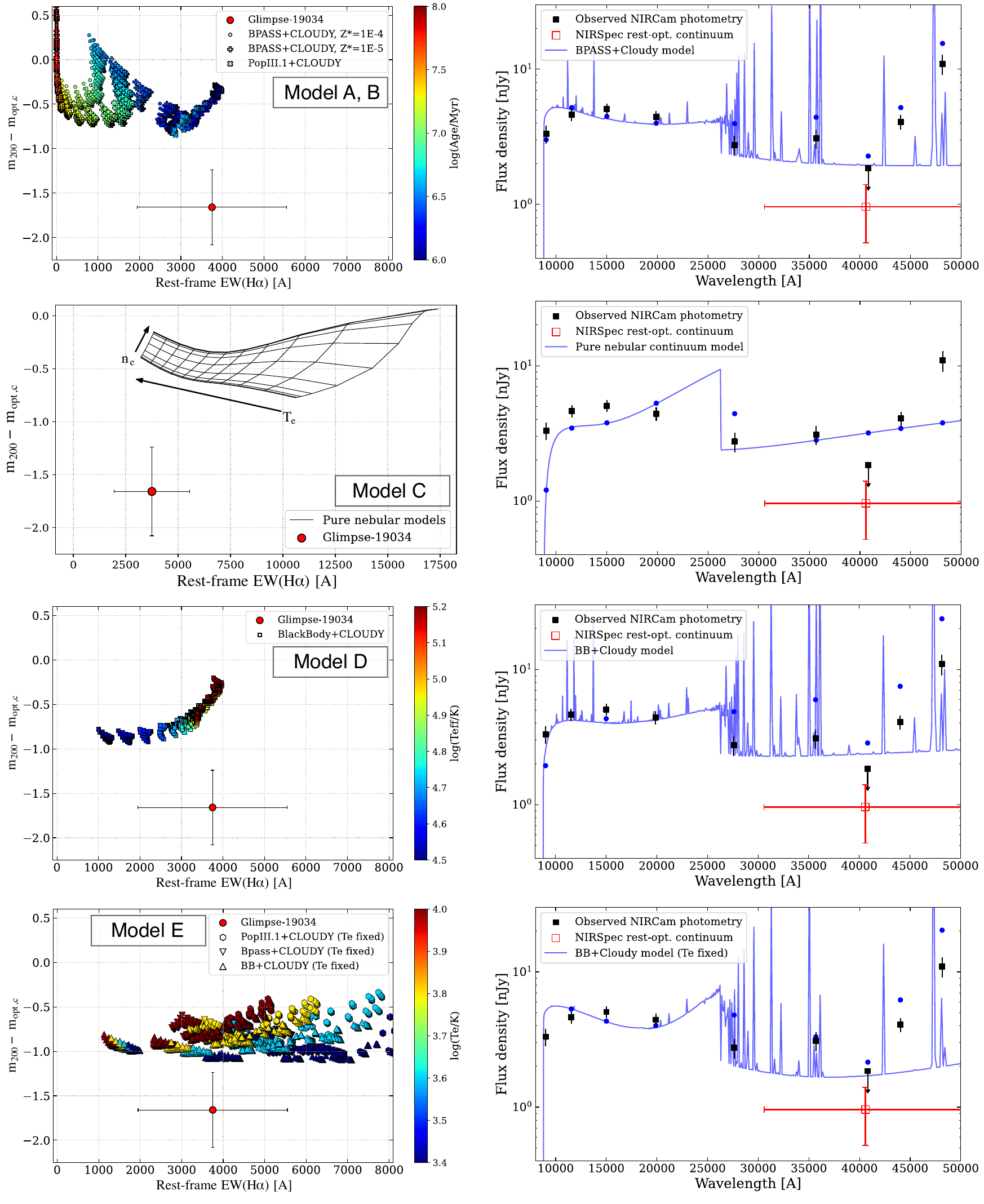}
\end{center}
\vspace{-0.4cm}
\caption{
Comparison of the observed properties of \targ\ with predictions from different photoionization models. 
The left panels show the parameter space of rest-frame EW(H$\alpha$) versus Balmer-jump amplitude ($m_{200}-m_{\rm opt,c}$) for each model setup: (A,B) BPASS/Pop~III$+$\textsc{Cloudy}, (C) pure nebular continua, (D) blackbody$+$\textsc{Cloudy}, and (E) \textsc{Cloudy} with fixed $T_{\rm e}$. 
The observed values for \targ\ are indicated with red symbols and uncertainties. 
The right panels present representative SED fits for each case, with the observed NIRCam photometry (black points) and the best-fit continuum level based on the G395M observations (red square), compared to the model spectra (blue curves). 
As summarized in Table~\ref{tab:model_summary}, none of the models simultaneously reproduce the extraordinarily strong Balmer jump and large EW(H$\alpha$) observed in \targ, reestabishing the persistent tension between current photoionization frameworks and the data.
\label{fig:cloudy}}
\end{figure*}

To investigate whether the observed strength of the H$\alpha$ line and Balmer jump in \targ\ can be explained within standard photoionization frameworks, we perform a series of modeling experiments considering both stellar+nebular synthesis and pure nebular emission scenarios. 
Each model is redshifted to $z_{\rm spec}=6.20$, and we compute NIRCam/F200W synthetic photometry and the continuum level at 4.1 $\mu$m to compute the Balmer jump amplitude of $m_{200} - m_{\rm opt,c}$ from the model, for the sake of fair comparison with observations of \targ.

\paragraph{(A) BPASS + \textsc{Cloudy} (stellar+nebular).}\,
We first adopt binary stellar populations from BPASS (including the most metal-poor sets) as ionizing sources, coupled with \textsc{Cloudy} nebular models spanning
$-2.4 \le \log(Z_{\rm gas}/Z_\odot) \le -1.0$ and $2.5 \le \log(n_{\rm e}/{\rm cm}^{-3}) \le 5.0$.
These models reproduce the \emph{qualitative} trends of increasing EW(H$\alpha$) and Balmer-jump amplitude with younger ages and higher ionization. 
However, the maximum EW(H$\alpha$) and Balmer-jump amplitude reach only $\simeq 3000$\,\AA\ and $\simeq -0.7$ mag, respectively, falling short of the observed values by $>2\sigma$.

\paragraph{(B) Pop~III + \textsc{Cloudy}.}\,
Pop~III ionizing spectra from \citet{zackrisson2011}, coupled to the same nebular grids, behave similarly to (A) in their Balmer-jump predictions, with two notable differences.  
At very young ages ($t_{\rm age} \sim 1$ Myr), these models can produce extremely large H$\alpha$ equivalent widths up to $\sim 4000$\,\AA, consistent with the measured EW range.  
However, even in these extreme cases, the Balmer jump remains far too weak. 
Moreover, Pop~III-like systems are inconsistent with the observed \oiii/H$\beta$ ratio, although Pop~III stars may reside in metal-enriched gas during a “self-polluted” phase following the first supernovae \citep{fwang2012, rusta2025}.

\paragraph{(C) Pure Nebular.}\,
To test whether nebular continua alone can account for the large Balmer jump, we compute the nebular continuum emission, composed of free–bound, free–free, and two-photon emission, on grids of $3.4 \leq \log(T_{\rm e}/{\rm K}) \leq 4.6$ and $2.0 \leq \log(n_{\rm e}/{\rm cm}^{-3}) \leq 7.0$ using \texttt{PyNeb}. 
The models show stronger Balmer jumps with decreasing $T_{\rm e}$ and increasing $n_{\rm e}$, although the latter also produces redder rest-frame UV and optical continua. 
No combination of parameters yields the observed Balmer jump: the maximum amplitude reaches only $\sim -0.8$ mag, even under extreme assumptions suppressing two-photon, free–free, and helium continuum contributions. 
Furthermore, the resulting red UV continua are inconsistent with \targ, which exhibits a much bluer slope of $\beta = -2.34 \pm 0.36$.

\paragraph{(D) Blackbody + \textsc{Cloudy}.}\,
The pure-nebular results suggest that an additional component in the rest-frame UV may be required to reproduce both the Balmer jump amplitude and the observed UV continuum shape. 
We therefore experiment with idealized blackbody (BB) sources as ionizing inputs, peaking in the rest-frame UV ($\log T_{\rm eff}/{\rm K}\simeq 4.5$–5.2). 
Such blackbodies with $\log T_{\rm eff}/{\rm K}\simeq 4$–5 are observed in tidal-disruption event (TDE) sources in the local Universe \citep{gezari2021}, making this scenario not entirely unfeasible. 
These models behave similarly to the BPASS and Pop~III runs at very young ages ($\sim 1$ Myr): hotter BBs ($\sim 10^5$ K) enhance [O\,\textsc{iii}] but weaken the Balmer jump, while cooler BBs ($\sim 10^{4.7}$ K) yield colors closer to the observations but still fail to reproduce the extreme Balmer-jump amplitude, reaching at most $\sim -0.8$ mag.

\paragraph{(E) BPASS/Pop~III/BB + \textsc{Cloudy} with Fixed $T_{\rm e}$.}\,
Finally, we also run \textsc{Cloudy} models for all three ionizing sources (BPASS, Pop~III, BB) with the electron temperature fixed rather than solved self-consistently. 
While the models above adopt relatively simple assumptions, more complex conditions may exist in reality (e.g., a distribution of ISM properties, non-uniform geometries, or variations in the IMF). 
The $T_{\rm e}$–fixed experiments therefore provide a flexible test to probe a broader parameter space and to assess whether such effects could alleviate the discrepancies with the data. 
Among these runs, the BB+\textsc{Cloudy} model with $T_{\rm eff} \approx 10^{4.7}$ K and $T_{\rm e} \approx 6300$ K comes closest to the observed locus, yielding a Balmer-jump amplitude of $\sim -1.2$ mag. 
Although still insufficient, this value lies within $\sim1\sigma$ of the measurement. 
However, the corresponding \oiii/H$\beta \sim 1.0$ remains significantly lower than the observed ratio of $1.78 \pm 0.18$. 
While such a discrepancy might in principle be mitigated in a two-zone ISM configuration, where strong \oiii\ emission arises from compact young star-forming regions, our results indicate that even the flexible setups of (E) cannot simultaneously reproduce all of the observed properties of \targ\ without tension.

\begin{table*}[t!]
\setlength{\tabcolsep}{2pt}
\centering
\caption{
Comparison of the observed properties of \targ\ with the different models explored in this work. 
Symbols indicate whether each model can reproduce the observations: 
$\circ$ = consistent; 
$\triangle$ = marginal or partial agreement (i.e., within $\sim1\sigma$ uncertainty and/or physically explainable); 
$\times$ = inconsistent.
}
\vspace{-0.2cm}
\label{tab:model_summary}
\begin{tabular}{lcccc}
\hline\hline
Model & Significant Balmer jump & Strong EW(H$\alpha$) & [O\,\textsc{iii}]/H$\beta$ & Blue rest-UV continuum \\
 & $-1.66 \pm 0.47$~mag & $3750 \pm 1800$\,\AA & $1.78 \pm 0.18$ & $\beta = -2.34 \pm 0.36$ \\
\hline
(A) BPASS + \textsc{Cloudy}   & $\times$ ($\sim-0.7$) & $\triangle$ ($\lesssim3000{\rm \AA}$) & $\circ$ & $\circ$ \\
(B) Pop~III + \textsc{Cloudy} & $\times$ ($\sim-0.8$) & $\circ$ ($\lesssim4000{\rm \AA}$) & $\triangle^{\dagger}$ & $\circ$ \\
(C) Pure nebular & $\times$ ($\sim-0.8$) & $\circ$ ($\lesssim 8000{\rm \AA}$) & $\circ$ & $\times$ \\
(D) Blackbody (BB) + \textsc{Cloudy} & $\times$ ($\sim-0.8$) & $\circ$ ($\lesssim4000{\rm \AA}$) & $\circ$ & $\circ$ \\
(E) BPASS/PopIII/BB + \textsc{Cloudy} (fixed $T_{\rm e}$) & $\triangle$  ($\sim-1.2$) & $\circ$ ($\lesssim8000{\rm \AA}$) & $\triangle$ ($\sim1.0$) & $\circ$ \\
\hline
\end{tabular}
\tablecomments{
$\dagger$ This requires a situation where Pop~III stars reside in metal-enriched gas during a ``self-polluted'' phase, following the first supernovae \citep[e.g.,][]{fwang2012,rusta2025}.}
\end{table*}

\medskip

Table~\ref{tab:model_summary} summarizes these outcomes, highlighting the persistent tension between the models and the observed properties of \targ.
Figure~\ref{fig:cloudy} also summarize the parameter spaces (left panel) and the closest SED (right panel) to the observed Balmer jump and the EW(H$\alpha$) properties.  
Overall, none of the models considered here are able to reproduce the observed properties of \targ, in particular its unusually strong Balmer jump. 
This suggests that more exotic physical mechanisms may be at play in \targ.

Note that we do not directly observe the Balmer jump in the G395M spectrum, and in principle alternative scenarios may remain, in which a very blue continuum extends smoothly from the rest-UV into the rest-optical, with the observed emission lines simply superimposed, eliminating the need for any significant discontinuity at the Balmer edge. To assess this possibility, we additionally fit the NIRCam photometry with a simple power-law continuum combined with the emission-line fluxes measured from the G395M data. We find that matching the extremely faint rest-optical continuum would require a very steep spectral slope ($\beta\ll-2.9$). 
Although extremely UV-blue SEDs can arise from young stellar populations with high escape fractions and a similarly blue galaxy has been reported at $z=9.25$ ($\beta_{\rm UV}=-2.99\pm0.15$; \citealt{yanagisawa2025}), \targ\ exhibits strong nebular emission lines, making a simple stellar origin difficult to reconcile with the this power-law SED scenario. This may motivate us to explore other SEDs, such as accretion disk models. However, the required slope of $\beta\ll -2.9$ is significantly bluer than can be produced by standard geometrically thin, optically thick accretion disks, whose asymptotic slope is limited to $\beta\simeq -2.33$ \citep[e.g.,][]{fabian2025}. Moreover, the observed rest-UV photometry of \targ\ is inconsistent with any single power-law continuum. 
Consequently, such a smooth, jump-free continuum model is even less plausible than any of the physically motivated scenarios explored in cases (A)–(E). A detailed discussion of these power-law tests is provided in Appendix~\ref{sec:power-law}.

\subsection{Possible Scenarios}
\label{ref:plausible}

Although we find the difficulty with any model to reproduce the observed properties without any tensions in Section~\ref{sec:model}, the results still indicate that the $T_{\rm e}$–fixed BB+\textsc{Cloudy} model with $\log T_{\rm eff}\!\simeq\!4.7$~K provides the closest match to \targ, especially the significant Balmer jump feature. Here we discuss possible physical scenarios that could produce such a hot, blackbody-like continuum component. 

\paragraph{(i) Tidal-disruption event (TDE).}\,
A first possibility is that we are witnessing a TDE, in which a star passing too close to a massive black hole is torn apart by tidal forces. 
At $z=6.20$, the measured $M_{\rm UV}=-15.9$ of \targ\ \citep{fujimoto2025} corresponds to a rest-frame $L_{\nu}(1500{\rm \AA}) \simeq 6\times10^{27}$ erg s$^{-1}$ Hz$^{-1}$, or an integrated UV luminosity of $L_{\rm UV}\sim10^{42}$ erg s$^{-1}$. 
Adopting a bolometric correction of a factor of a few (appropriate for thermal TDE spectra; e.g., \citealt{gezari2021}), the implied bolometric luminosity is $L_{\rm bol}\sim {\rm few}\times10^{42}$ erg s$^{-1}$. 
This is $\sim$1--2 orders of magnitude fainter than the brightest local TDEs observed in the optical/UV ($L_{\rm bol}\sim10^{44}$ erg s$^{-1}$), but falls well within the observed distribution of lower-luminosity events \citep{vanvelzen2020}. 
It is also possible that the \jwst\ observations captured the source during a fading phase, after the initial peak luminosity had already subsided, which would naturally make the event appear fainter.
Interestingly, \targ\ shows a point-source morphology with an upper limit on the effective size of $<40$~pc \citep{fujimoto2025}. 
Such a high stellar surface density can promote the formation of an intermediate-mass black hole (IMBH) through runaway stellar collisions, 
and subsequently induce TDEs of stars \citep{rantala2025}. 
Therefore, the compact nature of \targ\ is in line with the TDE scenario. 

We note that reproducing the observed Balmer jump requires the single blackbody component to peak tightly in the rest-frame UV. 
If the peak were shifted to longer wavelengths, the Balmer jump would be suppressed due to the contributions from the tail of the blackbody, while shifting to shorter wavelengths would make the red nebular continuum appear in the rest-frame UV regime. 
The best-fitting case with $\log(T_{\rm eff}/{\rm K})\simeq4.7$ therefore represents a rather finely tuned solution. 
Remarkably, this temperature is in excellent agreement with the effective temperatures commonly observed in local TDEs, providing additional support for the TDE interpretation. 

We further examined archival \textit{HST}/WFC3 imaging taken in Oct.~2015--Apr.~2016 \citep{lotz2017}. 
The GLIMPSE observations were obtained in Sep.~2024, corresponding to a temporal separation of $\simeq9$~yr in the observer frame ($\simeq1.3$~yr in the rest frame). 
At the location of \targ, we measure the flux density of $11\pm18$ nJy in HST/F160W, compared to $5.0\pm0.5$ nJy in NIRCam/F150W. 
Although the HST data are too shallow to draw firm conclusions, the non-detection in \hst\ is consistent with the low luminosities implied by the \jwst\ observations and does not contradict the TDE scenario.

\paragraph{(ii) Microquasar.}\,
An alternative explanation is that the observed emission arises from a microquasar-like system, extending into the ``miniquasar'' regime \citep[e.g.,][]{khlen2005}, centered around an intermediate-mass black hole (IMBH). Microquasars are X-ray binaries in which a stellar-mass black hole accretes gas from a companion star, producing a luminous accretion disk and relativistic jets that resemble a scaled-down version of quasar activity.
In such stellar-mass black hole X-ray binaries, the intrinsic thin disk peaks in soft X-rays ($k_{\rm B}T_{\rm in}\sim1$~keV) and is well described by a multi-temperature disk rather than a single blackbody \citep[e.g.,][]{mcclintock2006}. 
Several of the physical mechanisms operating in these systems, including a radiatively efficient inner disk, high mass-accretion rates that generate a luminous disk corona, and the presence of dense, scattering-dominated outflows driven by super-Eddington accretion, can also occur in the case of an IMBH accreting at super-critical rates. These conditions can substantially boost the luminosity and produce an emergent UV/optical continuum dominated by reprocessed emission, effectively extending the classical microquasar phenomenology into the ``miniquasar'' regime.  
Moreover, in super-Eddington accretion states with optically thick disk winds \citep[e.g., SS~433;][]{Margon1984}, the reprocessed emission from the outflow can produce a quasi-thermal optical/UV component that is approximately blackbody-like with $T_{\rm eff}\sim(5$--$7)\times10^4$~K \citep{fabrika2004, poutanen2007}. 
This temperature range coincides with the best-fit BB+$T_{\rm e}$ model for \targ, suggesting that a microquasar-like source embedded in a dense gaseous envelope could also plausibly account for the observed Balmer jump. 
In this interpretation, the extreme equivalent width of H$\alpha$ would then arise from the combination of strong nebular recombination in the surrounding medium and a very weak underlying stellar continuum.

\medskip
In summary, both the TDE and microquasar scenarios can, in principle, supply the hot, blackbody-like continuum required to reproduce the significant Balmer jump SED shape observed in \targ.
While the available data do not allow us to distinguish between them, the faint TDE and the microquasar disk winds both provide plausible pathways. 
Future deep time-domain monitoring will be critical to test these scenarios.

\section{Revisiting Photometric Pop~III Search}
\label{sec:revisit}

\begin{figure*}[t!]
\begin{center}
\includegraphics[width=1.\textwidth]{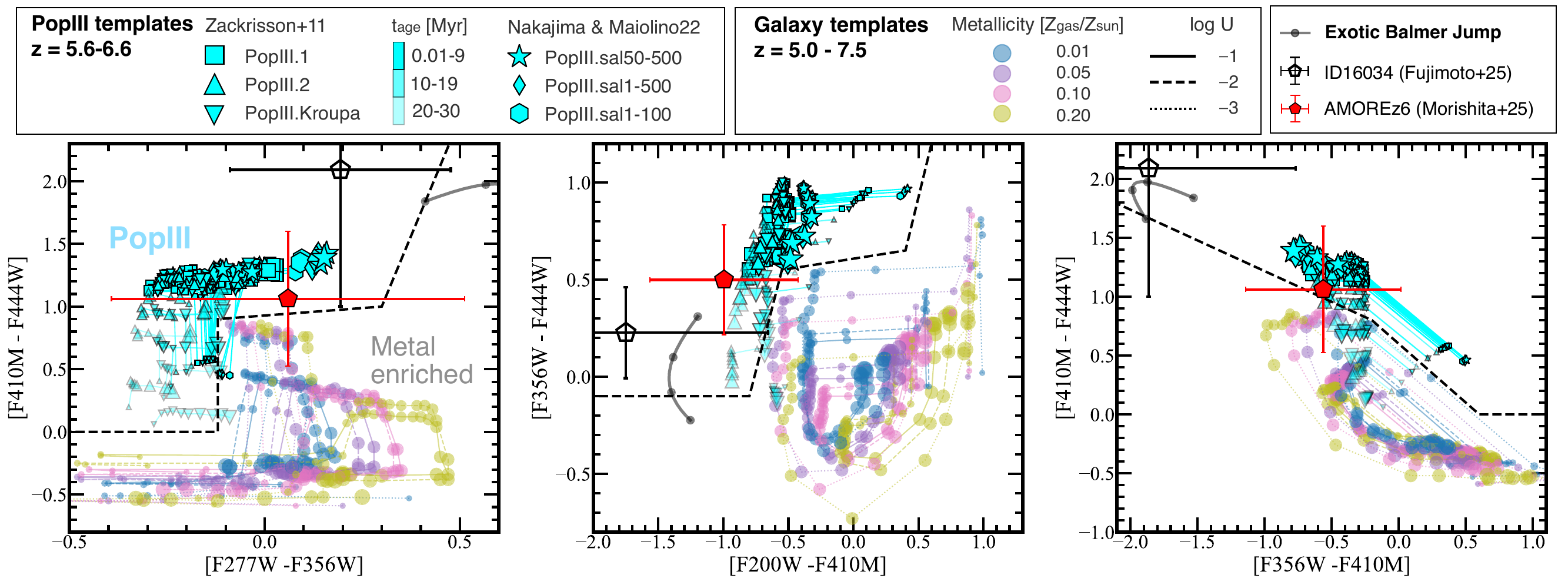}
\end{center}
\vspace{-0.4cm}
\caption{
NIRCam color--color diagrams for the Pop~III candidate selection at $z=5.6$--6.6, presented in \citet{fujimoto2025}.
The black dashed lines indicate the fiducial selection windows, designed to isolate promising Pop~III candidates exhibiting the key signatures of strong H$\alpha$, a prominent Balmer jump, and no obvious [O\,\textsc{iii}] emission. 
The cyan symbols show Pop~III SED model tracks at $z=5.6$--6.6 \citep{zackrisson2011,nakajima2022}, assuming the gas covering fraction of 1.0, while the colored circles indicate metal-enriched galaxy templates at $z=5$--7.5 generated with \texttt{Bagpipes} \citep{carnall2018}. 
For both Pop~III and metal-enriched galaxies, the symbol size decreases with increasing redshift, from $z=5.6$ to $z=6.6$ in steps of 0.1.
The open black pentagon marks \targ. 
The red pentagon marks the recently identified spectroscopic Pop~III candidate \targc\ \citep{morishita2025}, which is not included in the UNCOVER+Mega DR3 catalog due to blending with a nearby bright galaxy and was therefore missed in \citet{fujimoto2025}. 
The solid black curve traces the locus of exotic Balmer-jump objects (see Sec.~\ref{sec:selection}), which overlaps with the original Pop~III selection window and explains the contamination that occurred in the case of \targ.
\label{fig:pre_color}}
\end{figure*}

\begin{figure*}[t!]
\begin{center}
\includegraphics[width=1.0\textwidth]{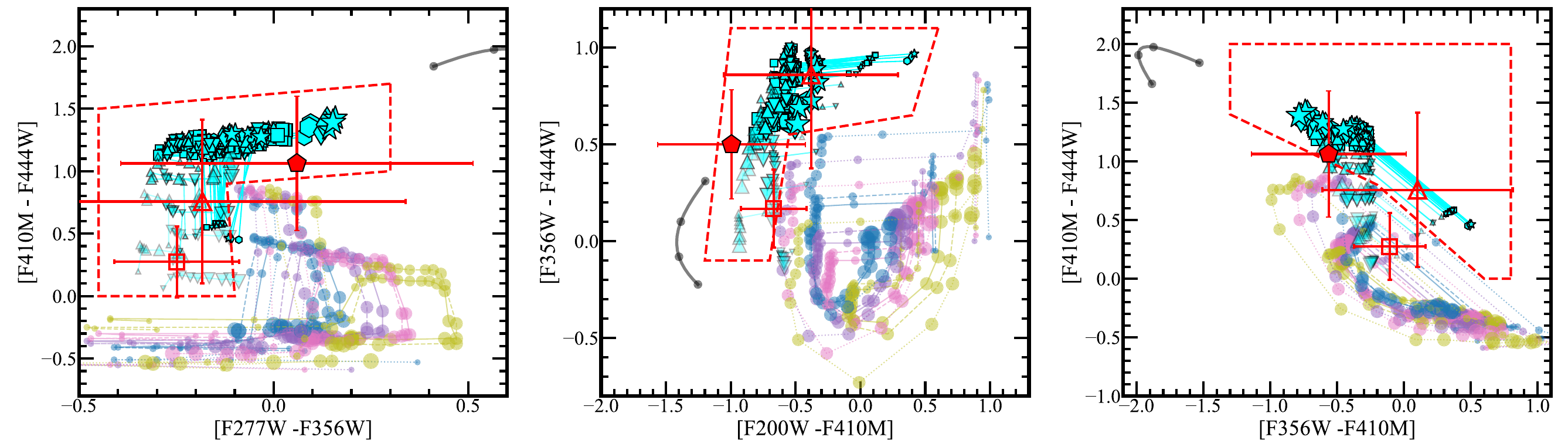}
\end{center}
\vspace{-0.4cm}
\caption{
Updated NIRCam color-color selection for Pop~III candidates at $z=5.6$--6.6. The symbols are the same as Figure~\ref{fig:pre_color}.
The red dashed lines indicate the revised color cuts, which cleanly separate the Pop~III model tracks (cyan symbols) from the color spaces of metal-enriched galaxies.
The revised selection avoids the locus of exotic Balmer-jump objects (black curve in Fig.~\ref{fig:pre_color}), while still recovering both simulated Pop~III tracks and the spectroscopic candidate \targc\ (red filled pentagon). 
The open red symbols represent tentative candidates found in strongly lensed LAEs: \targd c at $z=5.94$ (square) and \targe\ at $z=6.17$ (triangle) both identified in the Abell~370 field (see Appendix~\ref{sec:tentative}).  
\label{fig:new_color}}
\end{figure*}

Our and recent spectroscopic results motivate a reassessment of the photometric selection of Pop~III galaxy candidates. 
First, our NIRSpec follow-up of \targ\ has demonstrated that this source, originally identified by \cite{fujimoto2025} using the color- and SED-based selections, is not a bona-fide Pop~III system but instead an exotic galaxy with an extraordinarily strong Balmer jump. 
This case demonstrates that a subset of the original color space is vulnerable to contamination by such extreme Balmer-jump objects. 
Second, a new spectroscopic Pop~III candidate at $z=5.725$ has recently been reported in the Abell2744 field, \targc\ \citep{morishita2025}. This source lies within the UNCOVER footprint that was photometrically searched by \citet{fujimoto2025}. However, it turns out that \targc\ is not included in the UNCOVER+Mega DR3 catalog used in that analysis due to blending with nearby bright galaxies, remarking the caution for the possible incompleteness due to the chance projection with bright foreground galaxies. 

Using a $0\farcs2$-diameter aperture photometry for AMORE6-B on the NIRCam images with carefully subtractions of the foreground galaxies provided by \cite{morishita2025}\footnote{
The photometry is newly added in the revised version of \cite{morishita2025}, which is shared by T.~Morishita in private communication.
},  
we find that its photometry places the source well within the Pop~III color selection in all the three diagnostic planes originally presented in \cite{fujimoto2025}. We also find that its SED-based properties also satisfy the selection function, suggesting that \targc\ would have been identified as a promising candidate. 
These results emphasize the importance of ensuring that future photometric search are both robust against exotic contaminants, like \targ, and sufficiently inclusive so as not to miss promising Pop~III candidates like \targc.

Guided by these two lessons, in the following subsections we refine the photometric color selection (Sec.~\ref{sec:selection}), reapply the revised criteria to the legacy survey fields originally analyzed by \cite{fujimoto2025} and five new lensing cluster fields publicly released by the CANUCS team \citep{sarrouh2025} (Sec.~\ref{sec:research}), and update the inferred Pop~III UV luminosity function (UVLF) and Pop~III cosmic star formation rate density (CSFRD) (Sec.~\ref{sec:popiii_uvlf}). 

\subsection{Refining Selection}
\label{sec:selection}

\begin{figure*}[t!]
\begin{center}
\includegraphics[width=0.8\textwidth]{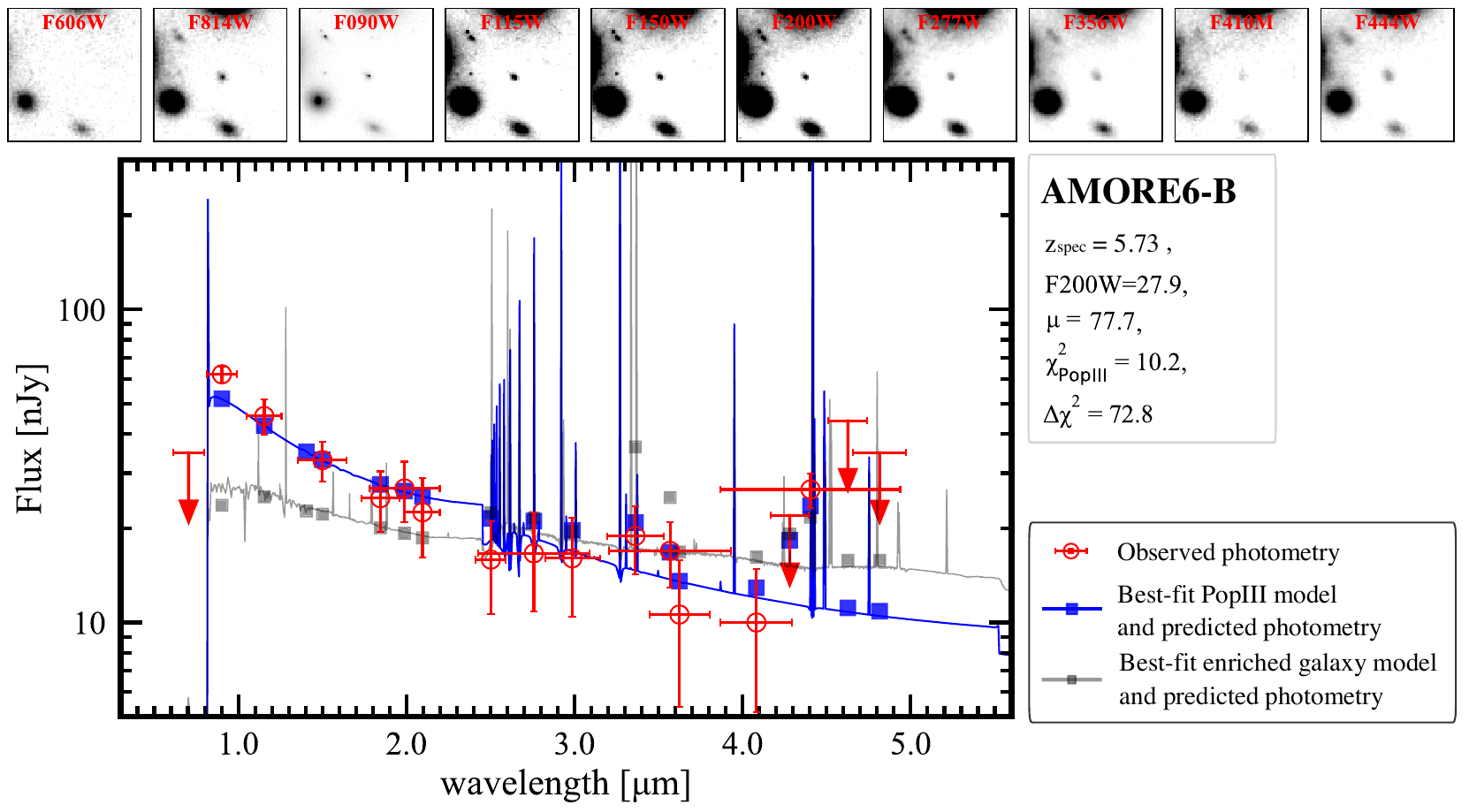}
\end{center}
\vspace{-0.4cm}
\caption{
Best-fit SED derived with \texttt{EAZY} using the Pop~III templates (blue curves) and the enriched-galaxy templates (grey curves) for the spectroscopic Pop~III candidates of \targc-B. 
For \targc-B, the absence of [O\,\textsc{iii}] emission is also confirmed independently in the NIRCam grism spectrum \citep{morishita2025}.
Postage-stamp $3''\times3''$ cutouts across all NIRCam filters are shown above each SED panel. 
Observed photometry (uncorrected for lensing) is displayed as red circles, while the blue and black squares indicate the model-predicted photometry in the HST and NIRCam filters for the best-fit Pop~III and enriched-galaxy solutions, respectively. 
Given the uncertainty of the total flux correction for strongly lensed, highly distorted sources, and possible source blending from nearby soureces,  we adopt the $0\farcs2$-diameter aperture-based photometry provided by \cite{morishita2025}.
}
\label{fig:spec_cand}
\vspace{0.2cm}
\end{figure*}

Figure~\ref{fig:pre_color} shows the original color--color diagrams presented in \citet{fujimoto2025}, along with the positions of \targ\ and \targc. 
The fiducial Pop~III selection was defined by the black dashed regions and designed to maximize purity by isolating sources with strong H$\alpha$ and Balmer-jump signatures while minimizing [O\,\textsc{iii}] contamination. 
Our spectroscopic follow-up, however, has demonstrated that exotic Balmer-jump galaxies such as \targ\ can also fall within this space, producing a clear contamination pathway. 
To illustrate this, the black curve in Figure~\ref{fig:pre_color} traces the locus occupied by such systems. 
It is constructed in the same manner as in Figure~\ref{fig:cont-oiiihb}: we vary the [O\,\textsc{iii}]/H$\beta$ ratio over the range 1.0--2.5, where the best-fit continuum becomes extremely faint and therefore produces a strongly Balmer-jump–like SED. 
For each value, we compute the corresponding continuum level together with the H$\alpha$ (+H$\beta$ and [O\,\textsc{iii}]) emission, and then convolve the resulting spectra with the NIRCam filter response curves to generate synthetic photometry.

To address these issues, we revise the photometric selection by explicitly excluding the region traced by extreme Balmer-jump objects. 
The updated selection windows, shown by the red dashed lines in Figure~\ref{fig:new_color}, remain consistent with the simulated Pop~III color distributions (cyan points) while avoiding the locus of Balmer-jump contaminants. 
Formally, the revised selection is defined by the following polygons:
\begin{itemize}
    \item F277W $-$ F356W vs.\ F410M $-$ F444W:
    \begin{equation}
    \begin{aligned}
    (0.2, 1.6),\; (0.2, 1.0),\; (-0.12, 0.9), \\
    (-0.1, 0.0),\; (-0.5, 0.0),\; (-0.5, 1.6)
    \end{aligned}
    \end{equation}

    \item F200W $-$ F410M vs.\ F356W $-$ F444W:
    \begin{equation}
    \begin{aligned}
    (0.6, 1.1),\; (0.4, 0.65),\; (-0.55, 0.55), \\
    (-0.7, -0.1),\; (-1.2, -0.1),\; (-1.0, 1.1)
    \end{aligned}
    \end{equation}

    \item F277W $-$ F356W vs.\ F410M $-$ F444W:
    \begin{equation}
    \begin{aligned}
    (0.8,0.0),\; (0.6, 0.0),\; (-0.2, 0.8), \\
    (-1.3, -1.4),\; (-1.3, 2.0),\; (0.8, 2.0)
    \end{aligned}
    \end{equation}
\end{itemize}
With these revised cuts, we confirm that \targc\ (red filled pentagon) still falls well within the Pop~III regions, demonstrating that inclusion is preserved in the updated definition. 

In \cite{fujimoto2025}, the ultimate candidate was obtained by combining the color-based selection with the SED-based criterion. 
The latter was evaluated using \texttt{EAZY} \citep{brammer2008} with both metal-enriched and Pop~III template sets, and comparing the quality of the fits. 
Specifically, defining the chi-square values from the two template sets as $\chi^{2}_{\rm galaxy}$ and $\chi^{2}_{\rm popIII}$, and $\Delta\chi^{2}\equiv\chi^{2}_{\rm galaxy}-\chi^{2}_{\rm popIII}$, the selection required
\begin{equation}
\label{eq:sed1}
    \Delta \chi^{2} \geq 9 \;\land\; \chi^{2}_{\rm popIII} < 10,
\end{equation}
or
\begin{equation}
\label{eq:sed2}
    \Delta \chi^{2} \geq 30 \;\land\; \chi^{2}_{\rm popIII} < 20.
\end{equation}

However, the photometry may benefit from exceptionally high S/N owing to its strong lensing magnification. 
In such cases, even the best-fit Pop~III models yield high absolute chi-square values, while the relative preference could remain overwhelming. To accommodate this situation, we extend the original criteria by adding an additional alternative requirement: in addition to Equations \ref{eq:sed1} and \ref{eq:sed2}, a source is also accepted if 
\begin{equation}
\label{eq:sed3}
    \Delta \chi^{2} \geq 50,
\end{equation}
thus the SED-based criterion is satisfied if any of Equations \ref{eq:sed1}, \ref{eq:sed2}, or \ref{eq:sed3} holds. 
This ensures that highly magnified, high-S/N sources are not missed despite their intrinsically elevated $\chi^{2}_{\rm popIII}$ values. 

In addition, motivated by the lesson from \targ\ that emphasizes the importance of directly detecting the underlying rest-frame optical continuum, we impose an additional requirement,
\begin{equation}
\label{eq:sed4}
 \;\land\; {\rm S/N}_{\rm F410M} \geq 2, 
\end{equation}
ensuring that candidates are detected in the F410M band and validity of the flux excess measurements in F356W and F444W. 
Furthermore, because redshift reliability is a key component of the selection, we also require that the Pop~III-based photometric redshift from \texttt{EAZY} ($z_{\rm phot,PopIII}$) lies within $z=5.6$--6.6 with a $1\sigma$ uncertainty of $\leq0.5$, and that the source remains undetected in the optical with ${\rm S/N}_{\rm F606W} < 2$. 
Together, these additional criteria restrict the selection to sources with consistent redshifts at $z=5.6$--6.6.

\begin{figure*}[t!]
\begin{center}
\includegraphics[width=1.\textwidth]{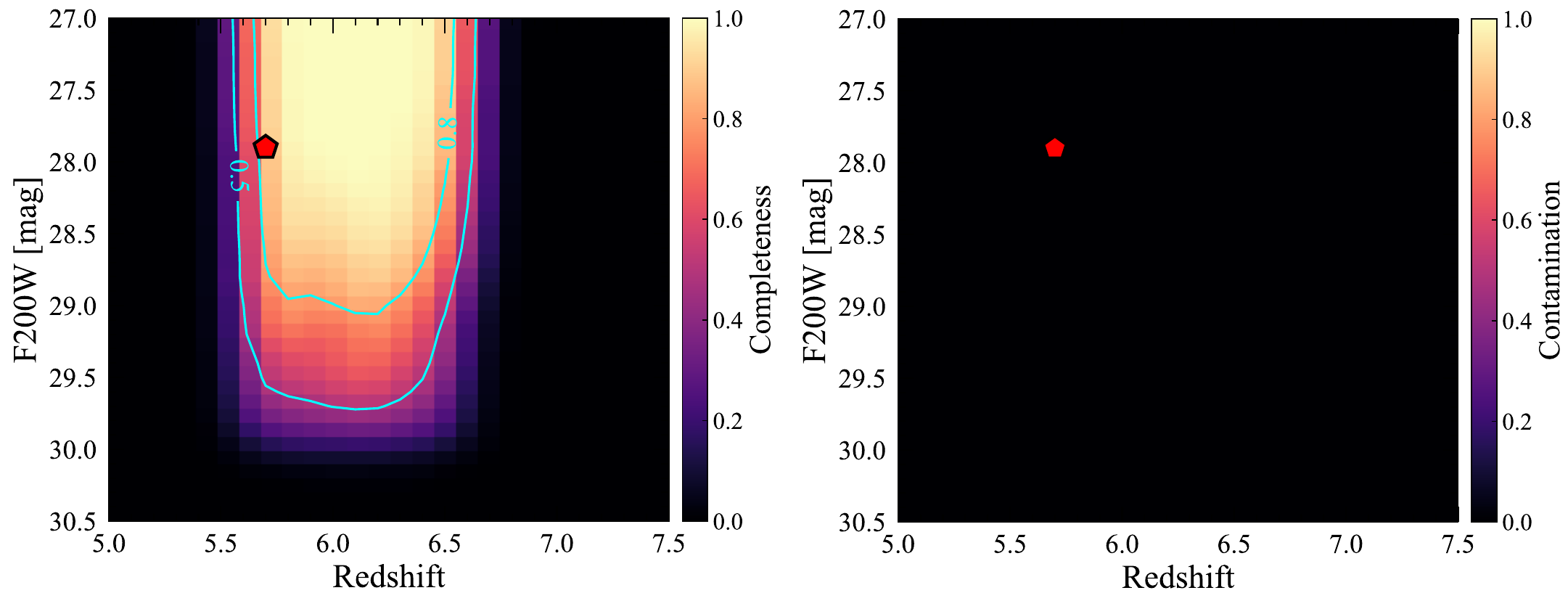}
\end{center}
\vspace{-0.3cm}
\caption{
Completeness (left) and contamination rate (right) distributions for our updated Pop~III photometric selection. 
The completeness is estimated by using Monte Carlo simulations with mock photometry is generated from Pop~III SEDs \citep{zackrisson2011}. 
The contamination rate is evaluated by using 1,472,791 simulated metal-enriched galaxies at $z=0-10$ from the Santa Cruz Semi-Analytic Model catalog \cite{yung2022}, with observational uncertainties applied. 
The red filled pentagon marks the spectroscopic candidate \targc\ \citep{morishita2025}, which is likewise confirmed to satisfy our updated NIRCam selection (Section~\ref{sec:research}).
The updated selection remains more restrictive than the original \citet{fujimoto2025} criteria, yet still achieves high recovery efficiency for Pop~III-like SEDs while substantially reducing the contamination risk from exotic Balmer-jump galaxies. 
}
\label{fig:comp-contami}
\end{figure*}

We re-evaluated the contamination rate and completeness of these updated color- and SED-based selections following the same Monte Carlo approach as in \cite{fujimoto2025}, generating mock photometry from simulated Pop~III SEDs (for completeness) and metal-enriched galaxies (for contamination rate) with observational uncertainties applied. 
In Figure~\ref{fig:comp-contami}, the resulting completeness and contamination rate distributions demonstrate that while the new selection is more restrictive, it retains high recovery efficiency for Pop~III-like SEDs while substantially reducing the contamination risk from exotic Balmer-jump galaxies. 

\subsection{Re-analysis for Legacy and New Fields}
\label{sec:research}

We next re-apply the updated color- and SED-based selections to the legacy survey fields initially analyzed in \citet{fujimoto2025} as well as to newly available datasets. 
In \citet{fujimoto2025}, the photometric search in the Abell~2744 field was based on the UNCOVER+Mega DR3 catalog. 
However, as discussed in Section~\ref{sec:selection}, this catalog missed the promising Pop~III candidate \targc\ due to blending with bright cluster galaxies. 
We therefore adopt the independent ALT catalog \citep{naidu2024}, which employs more aggressive source segmentation and successfully recovers \targc. 
For \targc, however, we use the dedicated aperture photometry measured on the NIRCam images with careful subtraction of the nearby foreground galaxies, provided by \citet{morishita2025}, given the possible flux contamination from the nearby galaxies. 
For the other fields analyzed in \citet{fujimoto2025}, we employed the same photometric catalogs as in the original study, namely: 
GLIMPSE (e.g., \citealt{atek2025}; see also \citealt{kokorev2025}), 
CEERS \citep[e.g.,][]{finkelstein2024}, 
PRIMER-UDS and PRIMER-COSMOS \citep[e.g.,][]{donnan2024},  
GOODS-S, including JADES \citep[e.g.,][]{eisenstein2023a, eisenstein2023b}, 
JADES Origin Fields (JOF; \citealt[e.g.,][]{robertson2023}), 
and FRESCO \citep[e.g.,][]{oesch2023}. 
This ensures consistency with the earlier analysis. 

In addition to revisiting the UNCOVER field, we also exploited the recently released CANUCS/Technicolor DR1 catalogs and lens models \citep{sarrouh2025}, which provide deep, homogeneous \jwst/NIRCam imaging of five massive lensing clusters (Abell~370, MACS0416, MACS0417, MACS1149, and MACS1423). 
These fields offer excellent depth ($\sim$29--30~mag) and lensing magnification, enabling sensitive searches for ultra-faint Pop~III candidates in more survey areas. 
We apply the revised color cuts (Figure~\ref{fig:new_color}) and the updated SED-based criteria (Equations~\ref{eq:sed1}--\ref{eq:sed4}) uniformly across the ALT, CANUCS, and the other legacy catalogs.

From this re-analysis, we find that no additional sources satisfy all of our selection criteria as robust Pop~III candidates beyond \targc. 
However, we identify four \textit{tentative} candidates that meet most of the requirements but fail either one of the updated color--color criteria or the new S/N threshold in F410M (Eq.~\ref{eq:sed4}). 
Interestingly, two of these four tentative candidates reside in known strongly lensed ($\mu\simeq7-15$) LAEs at $z=5.94$ and $z=6.17$ in Abell~370 \citep{ade2022}. 
The nature of these strong LAEs is reminiscent of recently reported spectroscopic Pop~III candidates at $z=3$--6 \citep{cai2025,morishita2025,vanzella2025}. 
Moreover, one of the two is identified as a compact stellar clump within a strongly lensed arc, where other regions in the arc exhibit strong [O\,\textsc{iii}] emission, suggesting the presence of a metal-poor pocket embedded within an otherwise enriched galaxy and highlighting the power of spatially resolved studies enabled by strong lensing.
Although we do not include these tentative sources in the following analyses, their colors are shown in Figure~\ref{fig:new_color} for reference, and further details for the spec-$z$ confirmed, two candidates identified in LAEs are provided in Appendix~\ref{sec:tentative}. 
We note that another tentative candidate initially reported in \citet{fujimoto2025}, JOF-21739, fails two of the three updated color--color criteria and is therefore even less prominent Pop~III candidate compared to the other tentative candidates.

For \targc, the colors are shown in Figure~\ref{fig:new_color}, and the best-fit SED is presented in Figure~\ref{fig:spec_cand}. 
Although deep spectroscopic follow-up is required to confirm the complete absence of the metal emission lines for for \targc, this outcome highlights the effectiveness of the refined selection in excluding contaminants such as exotic Balmer-jump galaxies while preserving promising Pop~III candidates.

\subsection{Updated Pop~III UVLF and CSFRD}
\label{sec:popiii_uvlf}

\begin{figure*}[t!]
\begin{center}
\includegraphics[width=0.75\textwidth]{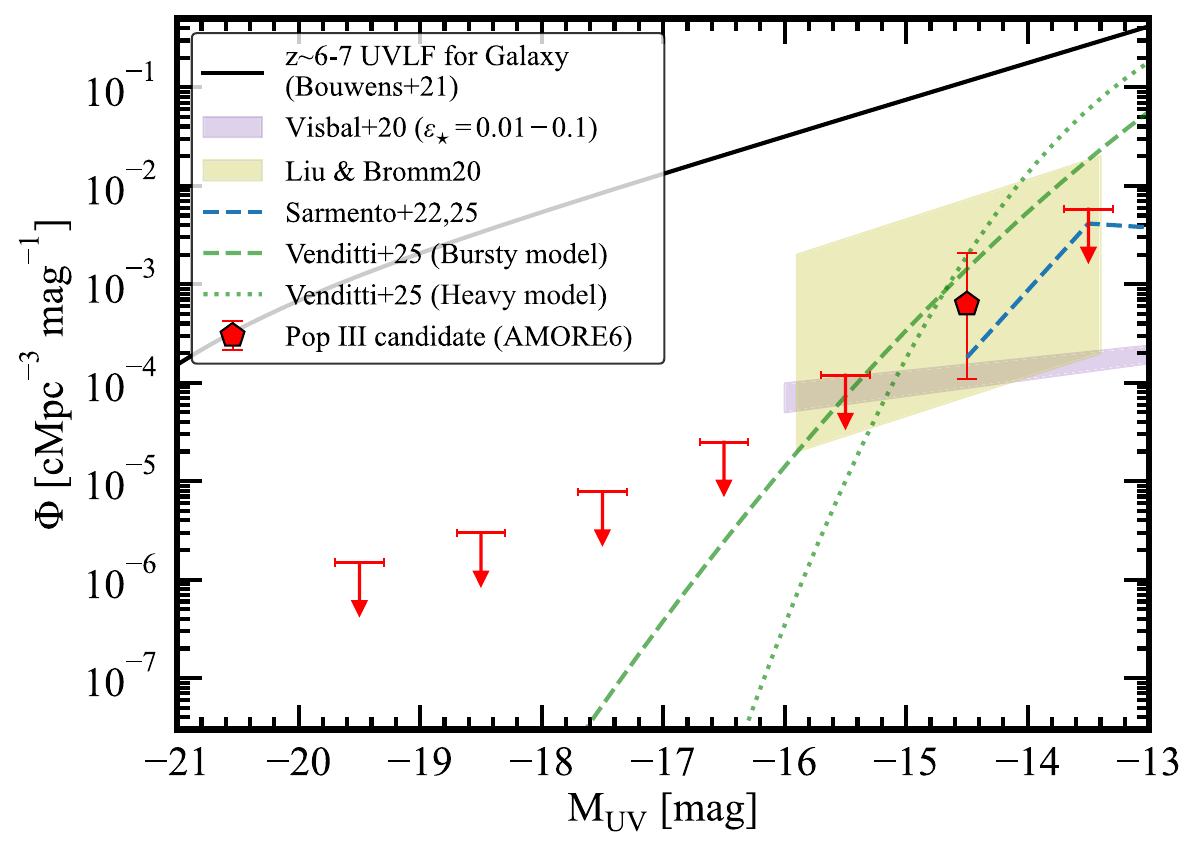}
\end{center}
\vspace{-0.3cm}
\caption{
Updated Pop~III UV luminosity function (UVLF) at $z \simeq 6$--7. 
The red point shows AMORE6, and arrows indicate the 1$\sigma$ upper limits based on the Poisson uncertainty presented in \cite{gehrels1986}. 
The black curve is the UVLF of general $z \simeq 6$--7 galaxies \citep{bouwens2021}, and shaded regions and curves show theoretical Pop~III predictions \citep[e.g.,][]{visbal2020, liu2020, sarmento2022, venditti2025}. 
}
\label{fig:uvlf}
\end{figure*}

Given the recent example of LAP1-B, initially identified as a Pop~III candidate due to the apparent absence of [O\,\textsc{iii}] in its NIRSpec spectrum \citep{vanzella2023}, but later revealed to exhibit multiple metal lines in deeper spectroscopy \citep{nakajima2025}, deep follow-up observations will be essential to first definitively establish whether \targc\ is truly a Pop~III system. Nevertheless, for the purpose of exploring the observational implications, we proceed under the assumption that \targc\ is a Pop~III galaxy and revisit the resulting constraints on the Pop~III UV luminosity function (UVLF) and cosmic star-formation rate density (CSFRD) at $z\simeq6$--7.
We adopt the same methodology as in \citet{fujimoto2025}, combining the magnification correction for the survey volume, completeness correction, and Poisson statistics to derive the volume density of the Pop~III galaxies. 
Our reanalysis expands the survey volume by including the five CANUCS cluster fields \citep{sarrouh2025} in addition to the legacy datasets. 
We summarize the effective survey areas in Appendix~\ref{sec:survey_area}. 

Figure~\ref{fig:uvlf} presents the updated Pop~III UVLF. 
The red filled pentagon denotes the spectroscopic candidate \targc, and the red arrows indicate the $1\sigma$ upper limits in the remaining magnitude bins. 
Our constraints, including the upper limits, remain broadly consistent with theoretical expectations from recent simulations \citep{visbal2020,liu2020,sarmento2022,venditti2023,venditti2025}. 
Future large lensing cluster surveys with \jwst\ (e.g., VENUS; \citealt{fujimoto2025_6882}) will help disentangle these possible models and provide a deeper understanding of the first episodes of star formation in the early universe.
The corresponding number densities in each magnitude bin are summarized in Table~\ref{tab:uvlf}.

\begin{table}
\vspace{0.2cm}
\caption{Updated constraints on Pop~III UVLF at $z=5.6$--6.6}
\label{tab:uvlf}
\centering
\begin{tabular}{ccc}
\hline
$M_{\rm UV}$ & $\Phi$ \, [10$^{-4}$ cMpc$^{-3}$ dex$^{-1}$] & $N$ \\
\hline
$-19.5$ & $<0.01$ & 0 \\
$-18.5$ &  $<0.03$ & 0 \\
$-17.5$ &  $<0.08$ & 0 \\
$-16.5$ &  $<0.25$ & 0 \\
$-15.5$ & $<1.19$ & 0 \\ 
$-14.5$ & $5.34_{-4.42}^{+12.28}$ & 1 \\
$-13.5$ &  $<58.03$ & 0 \\ 
\hline
\end{tabular}
\tablecomments{
The upper limit is estimated from the Poisson uncertainty at the single-sided confidence level of 84.13\% presented in \cite{gehrels1986}. 
}
\end{table}

\begin{figure*}[t!]
\begin{center}
\includegraphics[width=0.75\textwidth]{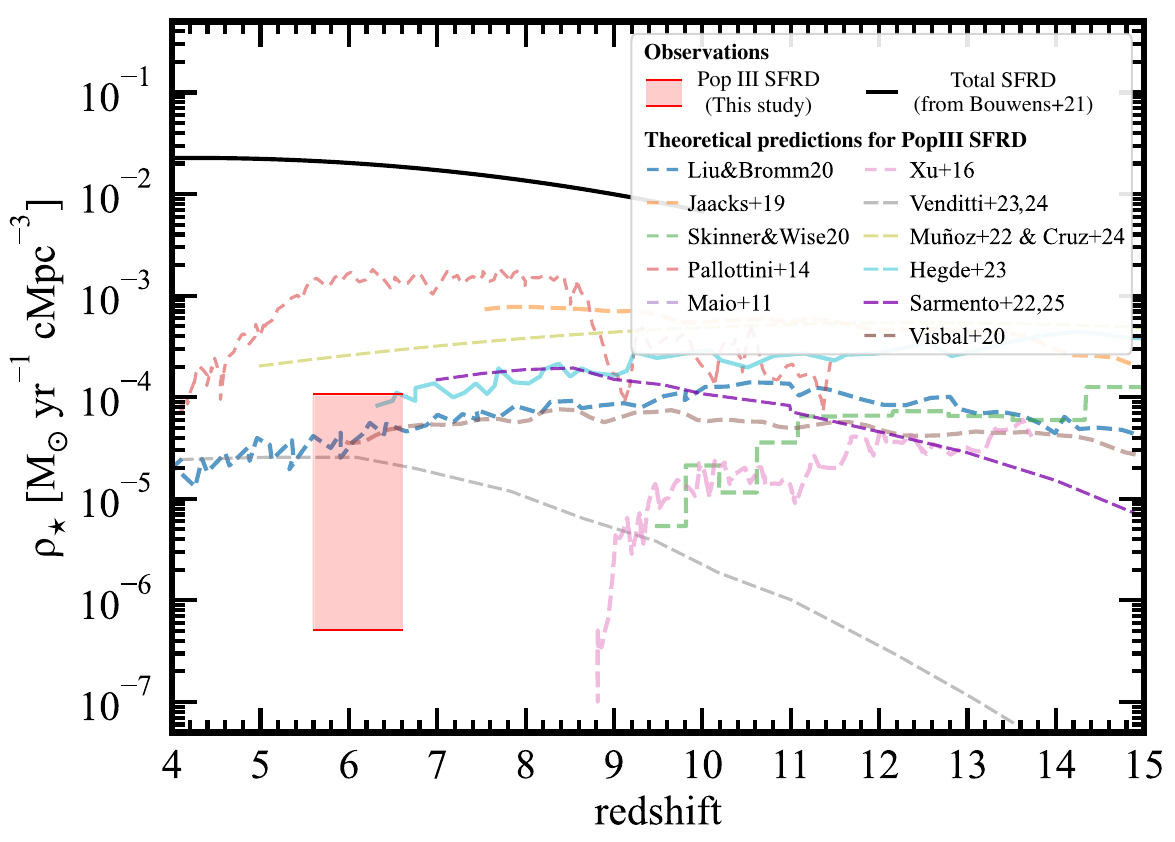}
\end{center}
\vspace{-0.3cm}
\caption{
Constraints on the cosmic Pop~III star-formation rate density (CSFRD) at $z\simeq6$--7. 
The red shaded region indicates our conservative lower and upper limits, derived under the assumption that AMORE6 is a genuine Pop~III system, and is compared to the total SFRD of galaxies (black curve; \citealt{bouwens2021}) and theoretical predictions for Pop~III SFRD (colored curves). 
This possible observational limit should be regarded as lower bounds, as our selection is optimized for young ($<10$~Myrs), Pop~III dominated systems with high nebular covering fractions, and may be incomplete for more complex or partially enriched populations. 
A fully self-consistent comparison with simulations that incorporate observational selection effects will be essential to fully interpret the observational implications for the cosmic contribution of Pop~III star formation.
}
\label{fig:csfrd}
\vspace{0.2cm}
\end{figure*}

\begin{table}[t!]
\caption{Updated constraints on Pop~III CSFRD at $z=5.6$--6.6}
\label{tab:csfrd}
\centering
\begin{tabular}{cc}
\hline\hline
Redshift & log($\rho_{\star,{\rm PopIII}}$) [M$_\odot$ yr$^{-1}$ Mpc$^{-3}$] \\
\hline
5.6--6.6 & [$-6.29$ : $-3.98$] \\
\hline
\end{tabular}
\end{table}

We also derive constraints on the Pop~III CSFRD by converting the UV luminosity density into a star-formation rate density using a Pop~III-specific UV conversion factor \citep[e.g.,][]{schaerer2002}. 
We derive the lower and upper limits in the same manner as \citet{fujimoto2025}: 
the lower limit is set by the detection of \targc\ alone, 
while the upper limit is obtained by rescaling the general galaxy UVLF to match the upper boundary of the data point by \targc\ and integrating it down to $M_{\rm UV}=-12$.

Figure~\ref{fig:csfrd} presents the updated CSFRD constraints, which are slightly lower than those in \citet{fujimoto2025}, reflecting the larger survey volume but the absence of additional detections. 
For comparison, we also show theoretical predictions (colored curves; \citealt{maio2011, pallottini2014, xu2016, jaacks2019, sarmento2022, skinner2020, visbal2020, liu2020, munoz2022, hegde2023, venditti2023}). 
Our updated limits remain broadly consistent with several theoretical predictions \citep[e.g.,][]{jaacks2019, liu2020, hegde2023}. They are also generally in line with the previous upper-limit estimate at $z\simeq4$–$5$ ($<5\times10^{-6},M_{\odot},\mathrm{yr}^{-1}$; \citealt{nagao2008}), which was based on searches for dual Ly$\alpha$ and \heii$\lambda1640$ emitters, although the level of incompleteness in that approach remains uncertain. Our updated limits on the Pop~III SFRD are summarized in Table~\ref{tab:csfrd}. 

We caution that our current search strategy is optimized for galaxies and star clusters dominated by Pop~III stellar populations, and may miss more complex systems. 
In particular, systems that have already been partially enriched by the first supernovae but still host surviving Pop~III star clusters are likely difficult to identify with our criteria \citep[e.g.,][]{fwang2012, rusta2025}. 
Similarly, galaxies with a low nebular covering fraction or not very young Pop~III star clusters ($>10$~Myrs) may fail to exhibit the very strong H$\alpha$ features required by our selection, and would thus be underrepresented. 
The first case may be revealed by spatially resolved studies of strongly lensed arcs, such as LAP1 \citep{vanzella2023, nakajima2025} and one of our tentative candidates found as a small clump in the strongly lensed arc (see Appendix \ref{sec:tentative}), which might represent metal-free pockets inside enriched systems. 
The ongoing large \jwst\ program VENUS (PID~6882; PI: S.~Fujimoto \& D.~Coe), which will establish homogeneous NIRCam 10-filter imaging for 60 massive lensing clusters, is expected to accelerate the discovery of such metal-free pockets and AMORE6-like pure Pop~III systems.
Regardless, future simulations that incorporate these observational constraints and adopt a self-consistent definition of Pop~III galaxies will be essential to fully interpret the observational limits and to trace the cosmic history of Pop~III star formation. 

\section{Summary}
\label{sec:summary}

In this paper, we present deep \jwst/NIRSpec G395M spectroscopy of GLIMPSE-16043, originally a promising Pop~III candidate reporeted in \cite{fujimoto2025} through \jwst/NIRCam photometry. 
Our main findings are as follows:

\begin{enumerate}
    \item We performed 30.4~hr of on-source NIRSpec/G395M observations of \targ. The spectra clearly detect H$\alpha$, H$\beta$, and [O\,\textsc{iii}]$\lambda\lambda4959,5007$, with line ratios of H$\alpha$/H$\beta=2.59\pm0.26$ and [O\,\textsc{iii}]/H$\beta=1.78\pm0.18$. These measurements show that \targ\ is not a genuine zero-metallicity Pop~III system.
    
    \item By combining the NIRSpec line fluxes with NIRCam photometry, 
    we constrain the underlying continuum and find an extraordinarily strong Balmer jump ($-1.66 \pm 0.47$ mag) and an extreme rest-frame H$\alpha$ equivalent width ($3750 \pm 1800$\,\AA). Both values exceed the predictions of standard stellar+nebular photoionization models 
    and explain why \targ\ was initially misidentified as a promising Pop~III candidate.

    \item We test a suite of physically motivated models to account for the extreme SED of \targ: (A) BPASS+Cloudy and (B) Pop~III+Cloudy reproduce some qualitative trends with the strong EW(H$\alpha$) but fail to match the Balmer jump; (C) pure nebular continua cannot simultaneously fit the Balmer jump and UV slope; (D) blackbody+Cloudy models with $T_{\rm eff}\sim10^{5}$ K enhance [O\,\textsc{iii}] but weaken the Balmer jump; and (E) blackbody+Cloudy models with fixed $T_{\rm e}$ provide the closest match with a $T_{\rm eff}\sim10^{4.7}$ K blackbody and $T_{\rm e}\sim6300$ K, yielding a Balmer jump amplitude of $\sim-1.2$ mag, still insufficient to reproduce all observed properties. This points toward more exotic origins.

    \item The origin of the single–blackbody, peaking in the rest-frame UV, may plausibly arise from either (i) a tidal-disruption event with $L_{\rm bol}\sim{\rm few}\times10^{42}$~erg~s$^{-1}$, or (ii) a microquasar-like system in which super-Eddington disk winds generate a pseudo-photosphere at $T\sim(5$--$7)\times10^{4}$~K. Both scenarios can, in principle, account for the best-fit blackbody component.

    \item These results demonstrate that photometric Pop~III color criteria are vulnerable to contamination by exotic Balmer-jump galaxies, especially when the rest-frame optical continuum is undetected. To mitigate this, we refine the \citet{fujimoto2025} selection by excluding the locus traced by such objects. The revised criteria also recover the recently reported spectroscopic candidate \targc\ \citep{morishita2025}, while excluding \targ. We emphasize, however, that the absence of \targc\ in the earlier search of \citet{fujimoto2025} was due to blending with nearby bright foreground galaxies, not a failure of the selection itself. This indicates that both the original and updated criteria maintain reasonable completeness, while the refined selection more effectively removes key contaminants.

    \item Reapplying the refined criteria to the \jwst\ legacy fields (UNCOVER, GLIMPSE, CEERS, PRIMER, JOF, JADES, FRESCO) and to the newly publicly released CANUCS cluster fields, we do not identify any additional sources that satisfy our Pop~III selection, beyond \targc. Still, we find four tentative candidates that meet most of the requirements but fail either one of the updated color--color criteria or the new S/N threshold in F410M. 
    Interestingly, two of these tentative candidates found in known strongly lensed LAEs at $z\simeq6$, reminiscent of recently reported spectroscopic Pop~III candidates at $z=3$--6 \citep{cai2025,morishita2025,vanzella2025}.
        
    \item Based on this re-analysis, and adopting AMORE6 as a Pop~III galaxy for the purpose of exploring the observational implications, we update the Pop~III UV luminosity function and cosmic star-formation rate density at $z\simeq6$--7. The larger survey area leads to constraints slightly lower than those reported in \citet{fujimoto2025}, yielding $\rho_{\star,{\rm PopIII}} \simeq [10^{-6}$--$10^{-4}]\,M_\odot\,{\rm yr^{-1}\,cMpc^{-3}}$, which falls within the range of current theoretical predictions. However, these values should be regarded as lower limits, since our selection is optimized for \emph{pure}, very young, high–nebular-covering-fraction Pop~III systems and is likely incomplete for more complex or partially enriched galaxies that may still host surviving Pop~III star clusters. A self-consistent definition of Pop~III populations, coupled with simulations that incorporate such observational selection effects, will be essential for fully interpreting the constraints on Pop~III star formation across cosmic time.
    \end{enumerate}
    
\acknowledgments{
We are grateful to Kohei Inayoshi and Pratika Dayal for helpful discussions for the physical interpretation of \targ, Adam Muzzin and Fengwu Sun for useful discussions, inputs, and sharing the data to investigate tentative candidates found in the Abell~370 field, and Takahiro Morishita for sharing the latest photometry of AMORE6-B.}

This work is based on observations made with the NASA/ESA/CSA James Webb Space Telescope. The data were obtained from the Mikulski Archive for Space Telescopes at the Space Telescope Science Institute, which is operated by the Association of Universities for Research in Astronomy, Inc., under NASA contract NAS 5-03127 for JWST. These observations are associated with program \#9223. Support for program \#9223 was provided by NASA through a grant from the Space Telescope Science Institute, which is operated by the Association of Universities for Research in Astronomy, Inc., under NASA contract NAS 5-03127.
The Dunlap Institute is funded through an endowment established by the David Dunlap family and the University of Toronto.
SF acknowledges support from the Dunlap Institute, funded through an endowment established by the David Dunlap family and the University of Toronto.
AV acknowledges funding from the Cosmic Frontier Center and the University of Texas at Austin’s College of Natural Sciences.

\bigskip

\appendix

\section{Updated photometry of GLIMPSE-16043}
\label{sec:new_photo}
The photometry used in \citet{fujimoto2025} was based on an early version of the data reduction and an internal catalog, whereas in this study we adopt the latest photometry for \targ. All updated fluxes remain within the $1\sigma$ uncertainties of the original measurements and do not alter the conclusions of \citet{fujimoto2025}. In fact, the new photometry even strengthens the previously reported photometric Pop~III signature: the quantity $\Delta\chi^{2}$ ($\equiv \chi^{2}_{\rm galaxy}-\chi^{2}_{\rm PopIII}$) increases when using the updated measurements. A comprehensive description of the revised imaging reduction and catalog construction is provided in the GLIMPSE survey release paper \citep{atek2025}; here we summarize the key updates relevant to this work.

Although the procedures for source detection, artifact removal, bCG subtraction, and catalog flagging follow the same principles as those used during the initial identification of GLIMPSE-16043 \citep{fujimoto2025}, a major update is the adoption of custom flat-fields. Specifically, we generated improved flat-field calibration frames using all publicly available NIRCam observations as of January 12, 2025. This is done due to the fact that a set of standard STScI supplied flat-fields introduced substantial correlated noise into the final mosaics, detrimentally affecting the effective depth of the images and introducing numerous spurious sources, especially in regions with where few dithers overlap. This problem was further exacerbated by the GLIMPSE observing strategy itself, which employed six wide dithers supplemented by four sub-pixel dithers at each position to optimize PSF sampling. While this pattern is ideal for PSF reconstruction, it also compounds flat-field residuals during co-addition. These residuals can resemble high signal-to-noise sources, particularly toward the mosaic edges where fewer large dithers overlap \citep[e.g. see the discussion in ][]{kokorev2025}. Implementing our custom flat-fields mitigates these effects and yields a measurable improvement in depth, and therefore S/N, reaching up to 0.5 magnitudes in the long-wavelength filters and approximately 0.2 magnitudes in the short-wavelength channels.

\begin{table}
\setlength{\tabcolsep}{30pt}
\begin{center}
\vspace{0.3cm}
\caption{NIRCam photometry of GLIMPSE-16043}
\label{tab:photometry}
\begin{tabular}{lc}
\hline 
\hline
R.A. [deg]  & 342.2123718      \\ 
Dec. [deg]  & $-$44.528751     \\ \hline
F090W [nJy] & $3.32 \pm 0.49$  \\
F115W [nJy] & $4.62 \pm 0.48$  \\
F150W [nJy] & $5.06 \pm 0.52$  \\
F200W [nJy] & $4.42 \pm 0.50$  \\
F277W [nJy] & $2.75 \pm 0.46$  \\
F356W [nJy] & $3.08 \pm 0.49$  \\
F410M [nJy] & $0.78 \pm 0.92$  \\
F444W [nJy] & $4.07 \pm 0.49$  \\
F480M [nJy] & $10.95 \pm 1.91$ \\ 
\hline
\end{tabular}
\end{center}
\vspace{-0.3cm}
\tablecomments{
The measurements above are aperture ($0\farcs3$-diameter) corrected photometry derived from PSF-matched images in the observed frame (i.e., without correcting for magnification).
}
\end{table}

\section{Power-law and emission-line model}
\label{sec:power-law}
In Section~\ref{sec:model}, we discuss that physically motivated stellar+nebular or pure photoionization models struggle to reproduce the observed SED properties of \targ. However, because the G395M 
spectrum does not directly detect a Balmer jump, we also test an extreme ``jump-free'' scenario in which a single, very blue power-law continuum extends smoothly from the rest-UV into the rest-optical wavelength. 
To assess this possibility, we fit the NIRCam photometry with a single power-law continuum,
\begin{equation}
    f_\lambda \propto \lambda^\beta,
\end{equation}
combined with the H$\alpha$, H$\beta$, and [O\,\textsc{iii}]\,$\lambda\lambda$4960,5008 fluxes measured from the G395M data. We set $\beta$ and the continuum normalization as free parameters.

\begin{figure}[t!]
\begin{center}
\includegraphics[width=0.5\textwidth]{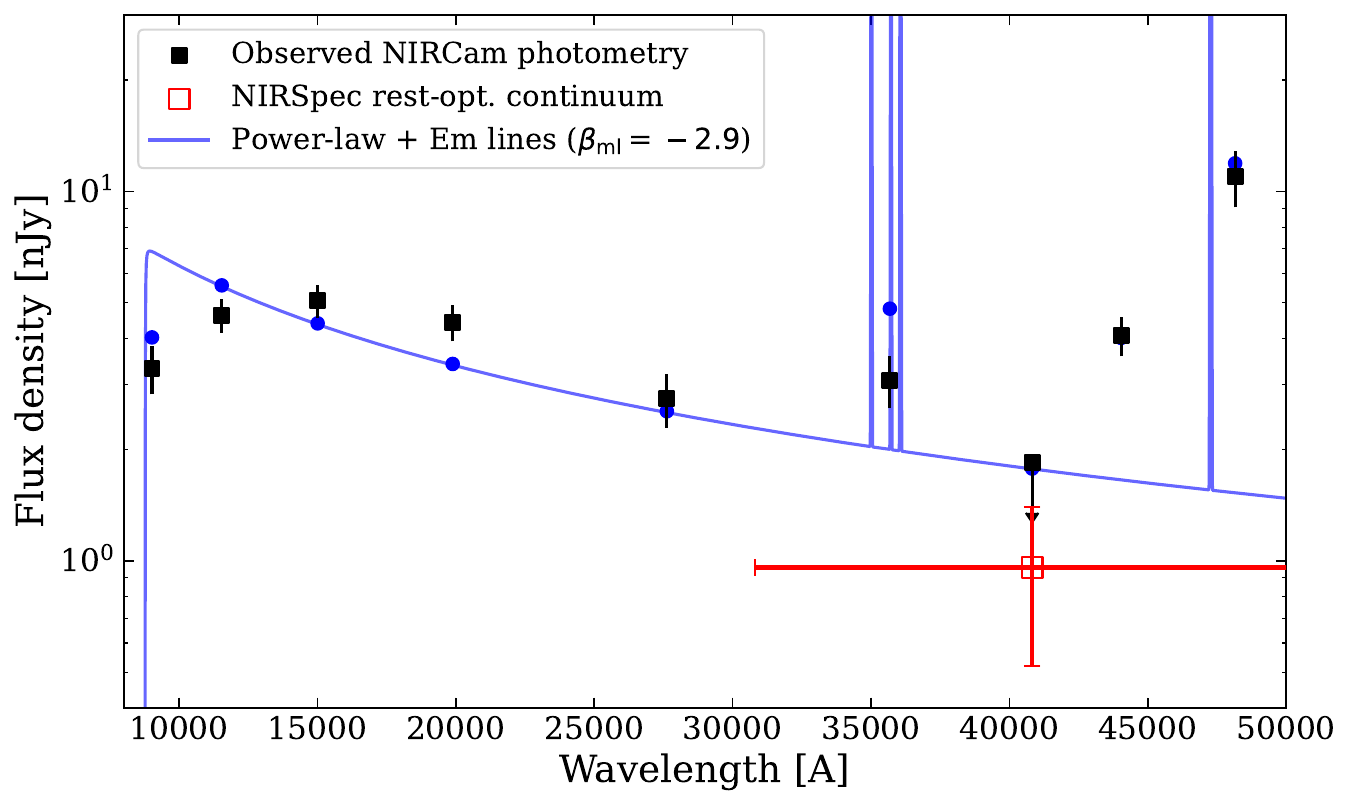}
\end{center}
\vspace{-0.4cm}
\caption{
Best-fit power-law $+$ emission-line model for \targ. The symobols are the same as Figure~\ref{fig:cloudy}. 
The best-fit model still underpredicts the G395M-based continuum estimate by $\sim$2$\sigma$ at $\sim$4$\mu$m and fails to match the observed rest-UV photometry, demonstrating that a smooth, jump-free continuum is also challenging to reproduce the SED of \targ.
}
\label{fig:powerlaw}
\end{figure}

In Figure~\ref{fig:powerlaw}, we present the best-fit power-law + emission line model. The best-fit model requires an extremely steep slope of $\beta_{\rm best} = -2.9$. Such extremely UV-blue SEDs can arise from young stellar populations with high escape fractions \citep[e.g.,][]{yanagisawa2025}, \targ\ exhibits strong nebular emission lines, making a simple stellar origin difficult to reconcile with the this power-law SED scenario. 
This motivates to other SED models, such as black hole accretion disk models \citep[e.g.,][]{tacchella2025,ji2025,fabian2025}, where the strong nebular emission may arise from surrounding narrow-line regions in the black hole accretion disk. 
Nevertheless, this is far bluer than physically achievable in standard, geometrically thin, optically thick accretion disks, whose power-law slope is analytically limited to $\beta \simeq -2.33$ ($f_\nu \propto \nu^{1/3} \Rightarrow f_\lambda \propto \lambda^{-7/3}$; e.g. \citealt{shakura1973}) and challenging to produce significantly bluer continua.
Even with $\beta = -2.9$, the model still underpredicts the rest-frame optical continuum inferred from our G395M-based continuum estimate by $\sim 2\sigma$ at $\sim$4~$\mu$m, implying that an even more extreme slope ($\beta < -2.9$) would be required well outside the realm of any plausible accretion-disk or radiative reprocessing model. Furthermore, the observed rest-UV photometry of \targ\ is incompatible with such a simple power-law shape, and the implied rest-frame EW(H$\alpha$) becomes even more extreme, $\gg 4000$\,\AA.  
These tests indicates that a smooth, jump-free continuum is even more challenging to reproduce the observed SED of \targ\ than the models discussed in Section~\ref{sec:model}.

\section{Tentative Candidates}
\label{sec:tentative}

\begin{figure*}[t!]
\begin{center}
\includegraphics[width=1.\textwidth]{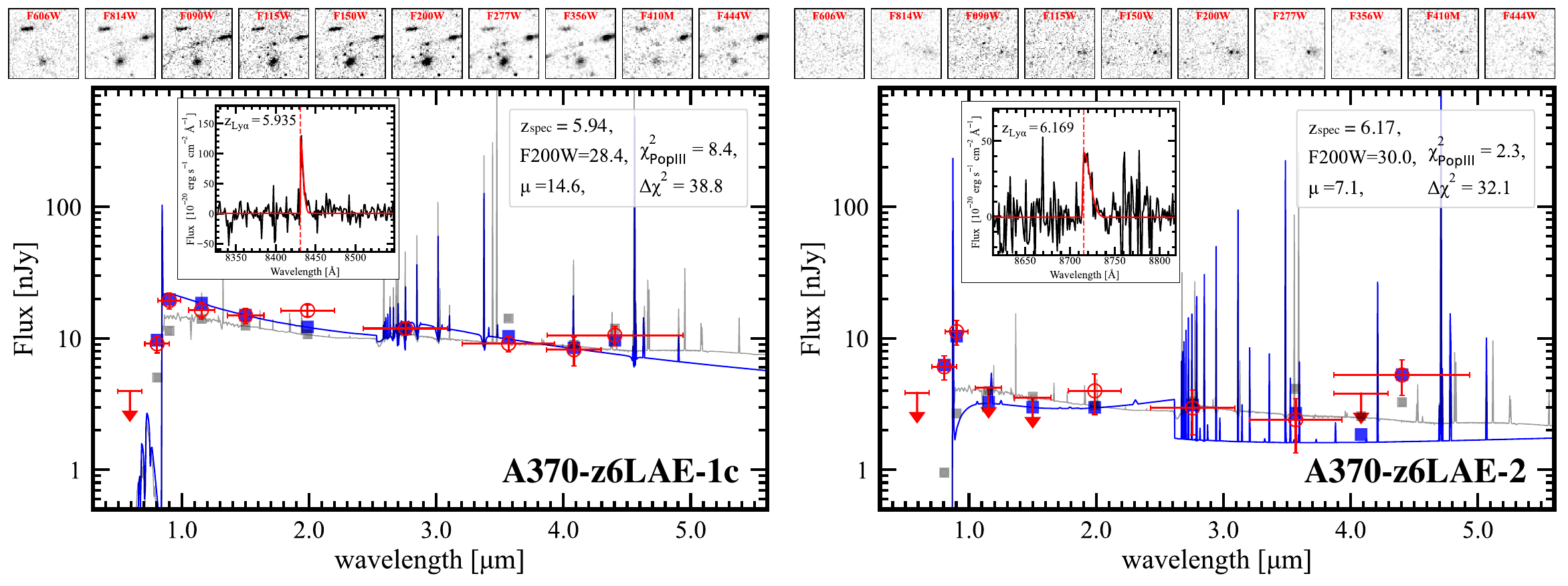}
\end{center}
\vspace{-0.3cm}
\caption{
Same as Figure~\ref{fig:spec_cand}, but for tentative candidates found in strongly lensed LAEs \citep{ade2022}: 
\targd c at $z=5.94$ (left) and \targe\ at $z=6.17$ (right), both located in Abell~370. 
\targd c is a compact stellar clump within the strongly lensed arc (see Figure~\ref{fig:z6lae1}). 
The inset panel shows the Ly$\alpha$ spectrum obtained with VLT/MUSE using a $0\farcs8$-diameter aperture centered on the NIRCam position.
The red curve represents the best-fit asymmetric Gaussian \citep[e.g.,][]{shibuya2014} to the Ly$\alpha$ line profile, and the red dashed vertical line denotes the the Ly$\alpha$ redshift. 
}
\label{fig:tentative}
\end{figure*}

In Section~\ref{sec:research}, our re-analysis in all public legacy \jwst\ fields both in general and lensing cluster fields show that no additional sources satisfy all of our updated selection criteria as robust Pop~III candidates at $z=5.6$--6.6, beyond \targc. 
However, we identify four tentative candidates that meet most of the requirements but fail either one of the updated color--color criteria or the new F410M S/N threshold (Eq.~\ref{eq:sed4}). 
Interestingly, two of these four sources reside in known strongly lensed LAEs, reminiscent of recently reported spectroscopic Pop~III candidates at $z=3$--6 \citep{cai2025,morishita2025,vanzella2025}. 
Below, we describe the two tentative candidates associated with the lensed LAEs. 
Their best-fit Pop~III and metal-enriched galaxy SED models are shown in Figure~\ref{fig:tentative}.

{\flushleft \bf A370-z6LAE-1c} -- 
is identified as a tentative PopIII candidate residing within a strongly lensed LAE, dubbed \targd, at $z = 5.94$ behind the Abell~370 cluster, where it appears as one of five resolved stellar clumps with a lensing magnification of $\mu \simeq 14.6$. 
This magnification implies an intrinsic absolute UV magnitude of $M_{\rm UV} = -15.2$, based on the observed F150W magnitude of 28.6~mag with a $0\farcs3$-diameter aperture.
The source coordinates are (RA, Dec) = (39.9805416, $-1.5668469$), with the source ID in the CANUCS DR1 catalog is 2123895 \citep{sarrouh2025}.
\targdc\ satisfies our updated Pop~III selection, including the SED-based criteria shown in Figure\ref{fig:tentative}, except for the third color–color diagram ([F356W$-$F410M] vs.\ [F410M$-$F444W]) in Figure~\ref{fig:new_color}. Its location in this diagram remains consistent with Pop~III model tracks, but corresponds to intermediate stellar ages ($t_{\rm age} \simeq 10$–30 Myr), where the strength of H$\alpha$ emission begins to diminish. As a result, the color space becomes degenerate with that of metal-enriched galaxies.
Therefore, although the position of \targdc\ does not rule out a Pop~III nature, it highlights the strong degeneracy between age and metallicity in cases where H$\alpha$ is weak, which is precisely the reason why our third color cut excludes the region where Pop~III SED tracks overlap with enriched-galaxy models. 
For this reason, we classify \targdc\ as a tentative candidate.

\begin{figure*}[t!]
\begin{center}
\includegraphics[width=0.9\textwidth]{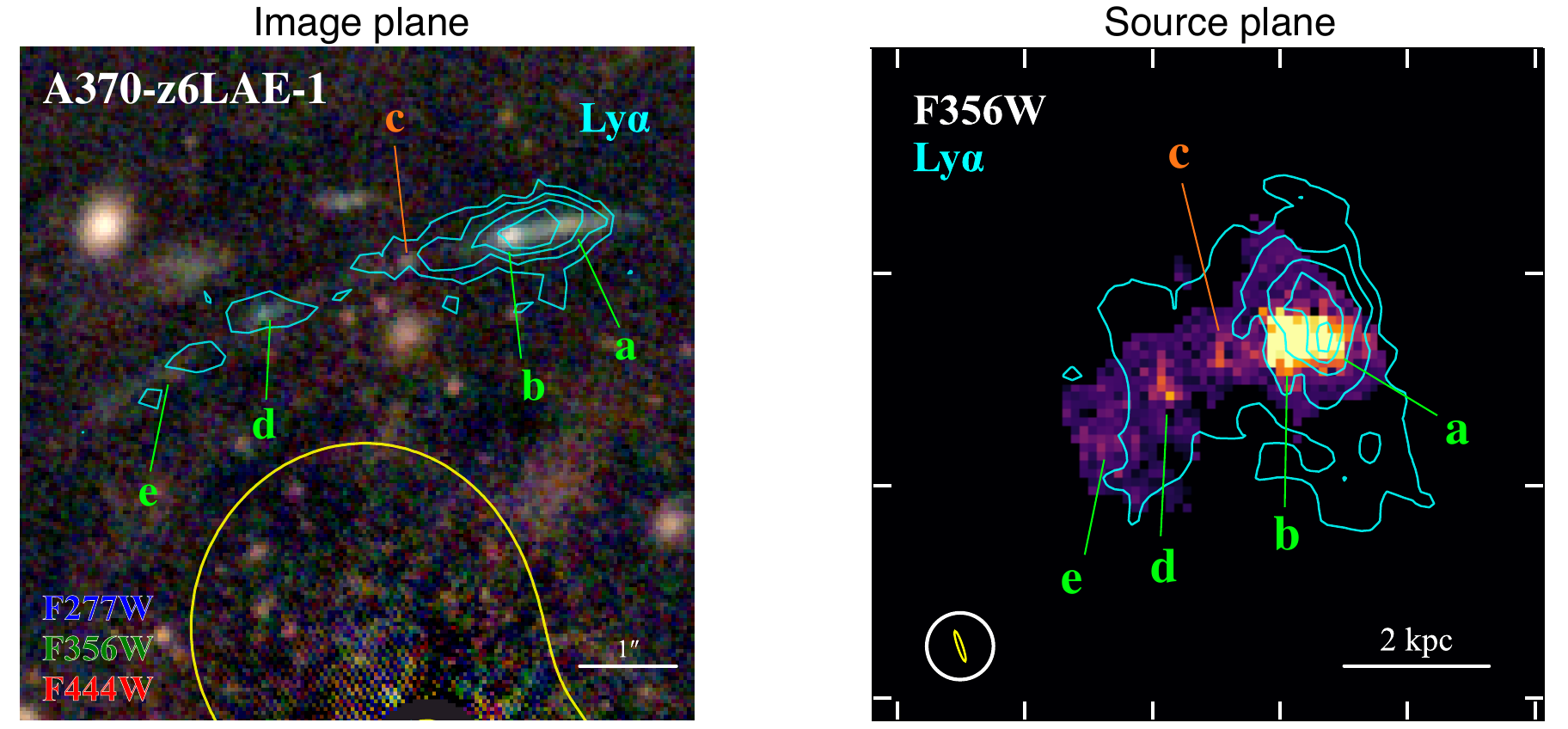}
\end{center}
\vspace{-0.4cm}
\caption{
Spatially resolved structure of \targd, illustrating the power of strong lensing for probing distinct and rare stellar populations such as Pop~III.  
\textit{\textbf{Left:}} NIRCam color composite image (R: F444W, G: F356W, B: F277W) of the strongly lensed system in the image plane. 
The cyan contours show the Ly$\alpha$ emission at $z=5.94$ (3$\sigma$, 5$\sigma$, 7$\sigma$, and 9$\sigma$) observed with VLT/MUSE \citep{ade2022}. 
The yellow curve represents the critical curve at $z=5.94$.
The five distinct clumps identified in the CANUCS catalog are labeled $a$–$e$, among which clump $c$ is selected as a tentative Pop~III candidate (see Appendix~\ref{sec:tentative}).
\textit{\textbf{Right:}} Source-plane reconstruction of the system in the F356W band, generated using the CANUCS Abell~370 lens model and the segmentation maps of clumps $a$–$e$. 
The yellow ellipse marks the effective PSF at the position of clump $c$ after lensing correction, while the white circle shows the nominal NIRCam/F356W PSF without lensing. 
The intrinsic morphology is highly elongated over $\sim$4~kpc, with clump $c$ situated between the bright core (clumps $a$ and $b$) and the tail region (clumps $d$ and $e$). 
The $\sim$1~kpc separation of clump $c$ from its neighbors demonstrates that Pop~III-like signatures in compact, metal-poor pockets can be captured with strong lensing support, yet would be smeared out at the nominal NIRCam resolution in general-field surveys.
}
\label{fig:z6lae1}
\end{figure*}

The left panel of Figure~\ref{fig:z6lae1} shows \targd\ in the image plane (i.e., observed frame without lensing correction), where the five clumps ($a$–$e$) are clearly resolved in the CANUCS NIRCam imaging \citep{sarrouh2025}, and the Ly$\alpha$ emission detected with VLT/MUSE \citep{ade2022} is spatially coincident with all of them.
The right panel presents the source-plane reconstruction of \targd, obtained using the latest publicly released CANUCS lens model for Abell370 \citep{sarrouh2025}. We adopt F356W for the reconstruction because this band provides the highest S/N for clump$c$ among the NIRCam long-wavelength (LW) filters. The host system exhibits an intrinsically elongated morphology spanning $\sim4$kpc, consisting of a bright core (clumps $a$ and $b$) and a tail hosting clumps $c$, $d$, and $e$. Clump$c$ lies between the brightest core clump~$b$ and the tail clump~$d$, with projected separations of $\sim$1~kpc from both.
This configuration is consistent with clump~$c$ being either a metal-poor pocket within the host galaxy of \targd\ or an infalling metal-poor satellite. Importantly, the $\sim1$kpc separation ($\simeq 0\farcs2$ at $z = 6$) from enriched neighbors demonstrates that PopIII-like signatures in compact metal-poor pockets can be easily diluted at the nominal NIRCam resolution in general-field surveys. This highlights the power of spatially resolved searches aided by strong lensing.

\begin{figure*}[t!]
\begin{center}
\includegraphics[width=1.\textwidth]{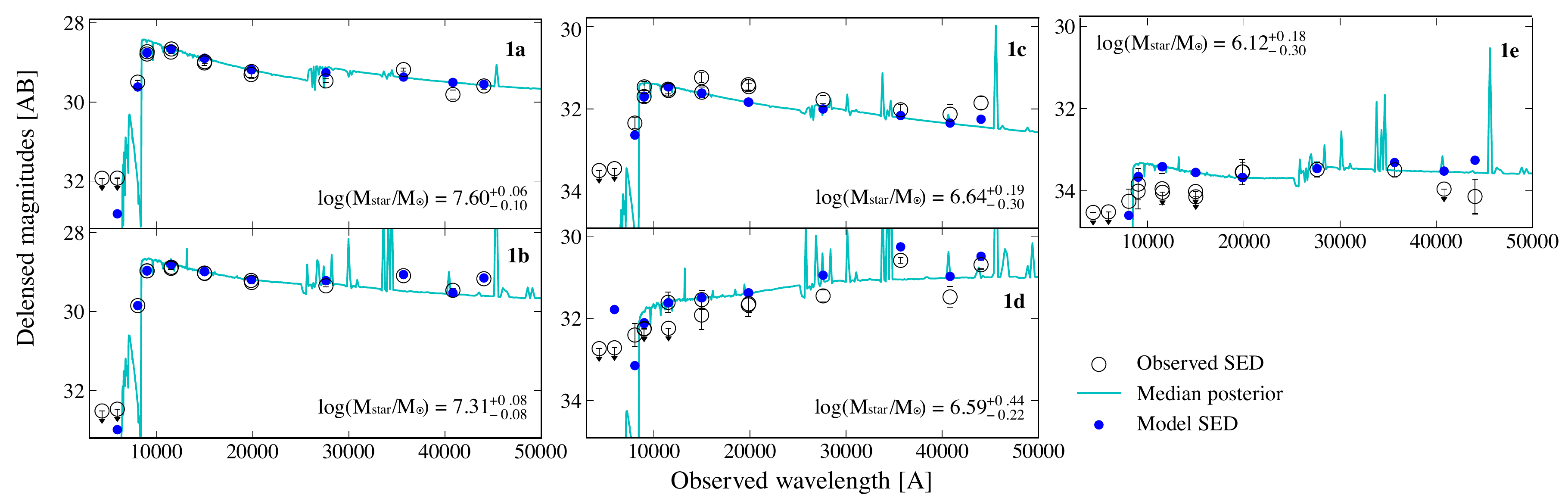}
\end{center}
\vspace{-0.5cm}
\caption{
Best-fit SEDs obtained with \texttt{Bagpipes} for each clump of \targd. The label indicates the inferred stellar mass for each clump. 
The black circles show the observed photometry (delensed), the blue points show the model-predicted photometry, and the cyan curves denote the median-posterior SEDs.  
The fits confirm that clump~$c$ exhibits very weak [O\,\textsc{iii}] emission, 
in contrast to the strong [O\,\textsc{iii}] seen in most of the other clumps. 
}
\vspace{0.5cm}
\label{fig:z6lae_sed}
\end{figure*}

To further characterize the system, we fix the Ly$\alpha$ redshift and flux and perform SED modeling with \texttt{Bagpipes} \citep{carnall2018} for each of the five clumps; their best-fit SEDs are shown in Figure~\ref{fig:z6lae_sed}.
The \texttt{Bagpipes} fits indicate that \targdc\ represents a low-mass ($\simeq10^{6.6},M_{\odot}$) stellar cluster hosted by a $\sim10^{8}\,M_{\odot}$ galaxy, and they confirm that clump~$c$ exhibits very weak [O\,\textsc{iii}] emission, in contrast to the strong [O\,\textsc{iii}] seen in most of the other clumps.
The weak [O\,\textsc{iii}] is also independently supported by the latest deep F360M imaging in this field (\#5890; PIs: C.~Withers \& A.~Muzzin), which shows no enhancement relative to F356W.
Conversely, the NIRCam F360M grism data obtained in the same region (PID~2883; PI: F.~Sun) show marginal positive flux at the expected \oiii$\lambda5008$ wavelength inferred from the Ly$\alpha$ redshift (integrated S/N $\sim$3), suggesting the possibility of non-zero [O\,\textsc{iii}] emission in \targdc. 
A key caution for both the imaging and grism measurements of \targdc\ is the presence of a nearby bright cluster-member galaxy, which makes the results highly sensitive to the details of local-background subtraction. 
Indeed, an independent reduction and analysis of the F360M grism data yield a weaker possible [O\,\textsc{iii}] feature with S/N $\lesssim2$, highlighting the level of practical uncertainty introduced by bright neighboring sources.
Taken together, these results reinforce the tentative nature of \targdc\ as a Pop~III candidate, while simultaneously demonstrating the natural importance of the spatially-resolved approach and how strong gravitational lensing enables us to probe distinct stellar populations inside a galaxy.

\begin{figure}[t!]
\begin{center}
\includegraphics[width=0.45\textwidth]{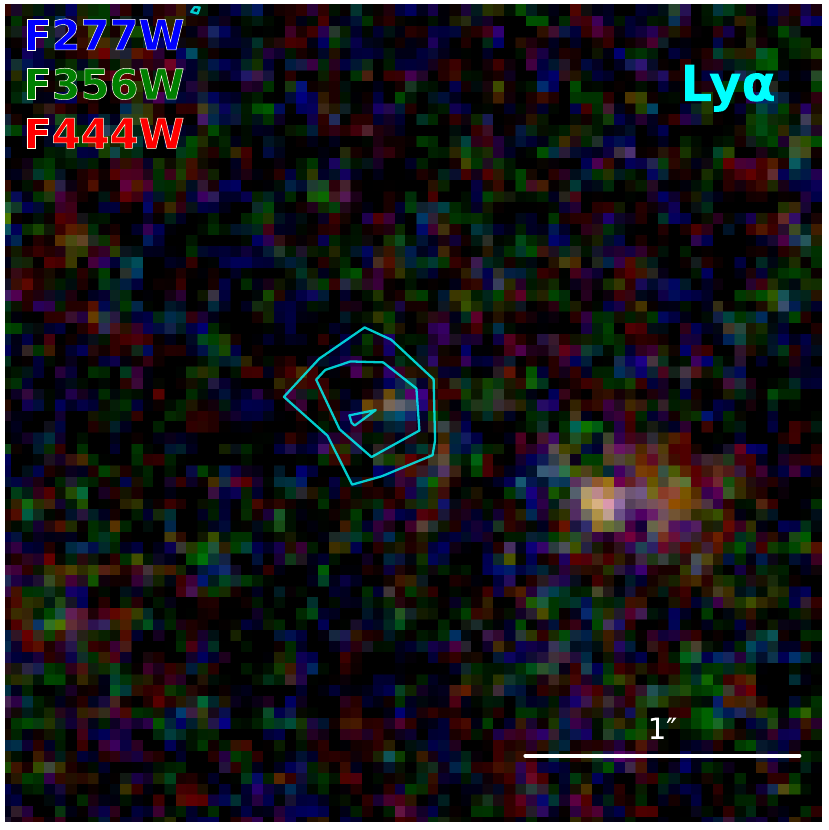}
\end{center}
\vspace{-0.3cm}
\caption{
Same as the left panel of Figure~\ref{fig:z6lae1}, but for another tentative candidate, \targe.
The cyan contours denote the Ly$\alpha$ emission at $z=6.17$ (3$\sigma$, 5$\sigma$, and 7$\sigma$) observed with VLT/MUSE \citep{ade2022}. 
}
\label{fig:z6lae2}
\end{figure}

{\flushleft \bf A370-z6LAE-2} -- 
is a compact LAE at $z = 6.17$ found behind the Abell370 cluster, with a lensing magnification of $\mu \simeq 7.1$.
Its observed F150W magnitude of 30.3~mag (1.7$\sigma$; measured in a $0\farcs3$-diameter aperture) implies an intrinsic absolute UV magnitude of $M_{\rm UV} = -14.3$ or even fainter.
The source coordinates are (RA, Dec) = (39.9814916, $-1.5658750$), with the source ID in the CANUCS DR1 catalog is 2114669 \citep{sarrouh2025}.
A NIRCam color composite of the system is shown in Figure~\ref{fig:z6lae2}.

\targe\ satisfies all of our updated PopIII selection criteria in both the color–color diagrams and the SED-based tests, except for the requirement of $\mathrm{S/N}_{\rm F410M} > 2$ (Eq.\ref{eq:sed4}).
The S/N in F410M is only $\sim$1.4, which naturally reflects the faintness of the object.
As discussed in Sec.~\ref{sec:selection}, this specific F410M requirement is designed to exclude contaminants exhibiting a significant Balmer jump, such as the case encountered for \targ.
Although the current position of \targe\ (red open triangle) in Figure~\ref{fig:new_color} lies well away from the region occupied by the exotic Balmer-jump contaminants, the low S/N in F410M permits its inferred location to scatter into that regime.
On the other hand, \targe\ is relatively isolated, with no bright neighboring sources, implying that deeper F410M imaging alone, even without the spectroscopy, could robustly establish its color and elevate it to a promising Pop~III candidate.

\section{Survey Area}
\label{sec:survey_area}

To compute the effective survey volume for identifying Pop~III candidates at $z\simeq5.6$--6.6, we follow the methodology of \citet{fujimoto2023b, fujimoto2025}, in which the source-plane area as a function of lensing magnification is evaluated for each cluster field. For the seven lensing clusters analyzed in this work -- five CANUCS clusters (Abell~370, MACS~0416, MACS~0417, MACS~1149, and MACS~1423), Abell~2744 from UNCOVER, and Abell~S1063 from GLIMPSE -- we use the latest publicly released lens models provided by the CANUCS \citep{sarrouh2025}, UNCOVER \citep{furtak2023a}, and GLIMPSE \citep{atek2025} teams. Because our $z\simeq5.6$--6.6 Pop~III color–color selection requires detections in all of F200W, F277W, F356W, F410M, and F444W, we restrict the survey area to regions where the full set of NIRCam bands overlaps. To this end, we manually define areas outlining the footprint in which all five filters are available at the required depth.

\begin{figure}[h!]
\begin{center}
\includegraphics[width=0.45\textwidth]{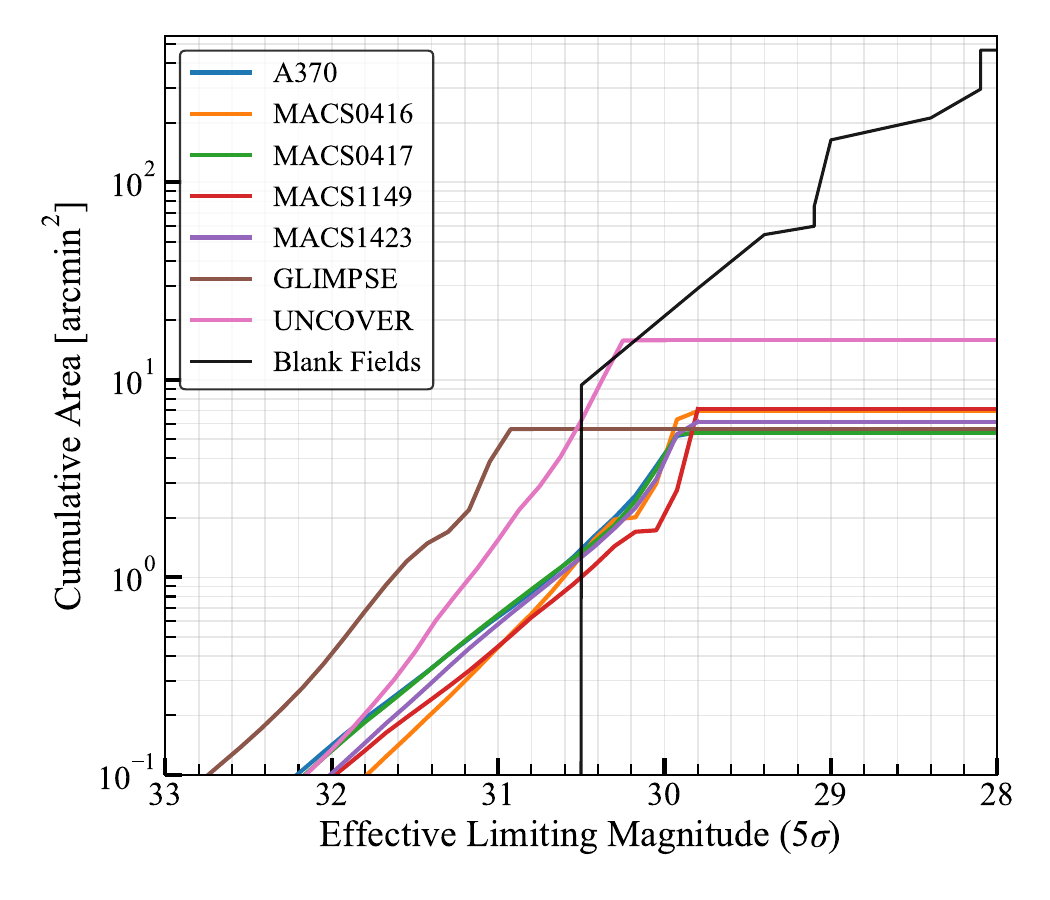}
\end{center}
\vspace{-0.4cm}
\caption{
Cumulative survey area as a function of the effective $5\sigma$ limiting magnitude for each field used in our Pop~III search. 
The colored curves represent the five CANUCS lensing clusters (Abell~370, MACS~0416, MACS~0417, MACS~1149, and MACS~1423), the GLIMPSE field (Abell~S1063), and the UNCOVER field (Abell~2744), computed using the latest available lens models. 
The black curve shows the corresponding area for non-lensed blank fields, derived from the limiting depths and survey areas compiled in \citet{fujimoto2025}. 
}
\label{fig:area}
\end{figure}

For each cluster, we compute the source-plane area as a function of magnification $\mu$, binned logarithmically from $\mu_{\rm min}=1$ to $\mu_{\rm max}=10^{3}$. At each pixel inside the defined region, the magnification is obtained from the lens model (assuming a source redshift of $z=6$), and the corresponding source-plane contribution is computed as the image-plane pixel area divided by $\mu$. Pixels are then accumulated into cumulative magnification bins, such that each bin records the total source-plane area with magnification exceeding a given threshold, $A(>\mu)$.
We then convert the relation between the cumulative source-plane area $A(>\mu)$ and $\mu$ into a relation between $A(>\mu)$ and the effective $5\sigma$ limiting magnitudes for each field. The limiting depths are obtained by scaling according to the magnification and by adopting the F200W $5\sigma$ sensitivities reported in the literature for the lensing fields analyzed in this study. For non-lensed general blank fields, we use the detection limits and survey areas compiled in \citet{fujimoto2025}. The resulting $A(>\mu)$--depth relations for all fields are presented in Figure~\ref{fig:area}. These relations are used directly to compute the magnification-corrected survey volume employed in our updated Pop~III UVLF and CSFRD analysis (Section~\ref{sec:popiii_uvlf}).

\bibliographystyle{apj}
\bibliography{apj-jour,reference}

\end{document}